\documentclass[aps,11pt]{revtex4}
\usepackage{dcolumn}
\usepackage{multirow} 
\usepackage{array}
\usepackage{makecell}
\usepackage{graphicx}
\usepackage{amsmath}
\usepackage{amsfonts}
\usepackage{amssymb}
\usepackage{psfrag}
\usepackage{wrapfig}
\usepackage{subfigure}
\usepackage{makeidx}
\usepackage{bm}
\usepackage{color}
\usepackage{epsf}
\usepackage[hypertexnames=false]{hyperref}
\hypersetup{hidelinks}
\usepackage{cases}
\usepackage{float}
\usepackage{placeins}
\usepackage{mathrsfs}

\makeatletter

\newcommand{\ud}{\,\mathrm{d}\,}

\newcommand{\Rmnum}[1]{\uppercase\expandafter{\romannumeral #1}}
\makeatother

\begin{document} 

\title{Time-domain analysis of scalar perturbations of Kerr--Newman black holes with Chern--Simons and Maxwell couplings}

\author{Yan-Ming Chen$^{1}$\footnote{yanmingchen\_mail@163.com},
Hyat Huang$^{1}$\footnote{hyat@mail.bnu.edu.cn},
Meng-Yun Lai$^{1}$\footnote{mengyunlai@jxnu.edu.cn}, De-Cheng Zou$^{1}$\footnote{Corresponding author: dczou@jxnu.edu.cn}, \\ Yun Soo Myung$^{2}$\footnote{ysmyung@inje.ac.kr} and Ming Zhang$^{3}$\footnote{ zhangming@xaau.edu.cn} }

\affiliation{$^{1}$School of Physics, Jiangxi Normal University, Nanchang 330022, China\\
$^{2}$Center for Quantum Spacetime, Sogang University, Seoul 04107, Republic of Korea\\
$^{3}$Faculty of Science, Xihang University, Xi'an 710077 China}

\date{\today}

\begin{abstract}
\indent
To study the onset of scalarization, we investigate linear scalar perturbations around Kerr--Newman black holes in the Einstein-Chern-Simons-Maxwell-scalar theory with quadratic Chern--Simons (CS) coupling $\gamma$ ($\hat\gamma=\gamma/M^2$) and Maxwell coupling $\eta$. For this purpose, we employ hyperboloidal R\'acz--T\'oth evolution and a stationary axisymmetric zero-mode computation. The resulting thresholds are compared with the time-domain response at representative points.
For $\hat\gamma=0$, we find two asymmetric threshold branches in $\eta$, while for $\hat\gamma=\eta$, the two branches have opposite signs and unequal magnitudes. For unequal CS and Maxwell couplings, the two real threshold branches merge as the spin or CS coupling increases.

\end{abstract}

\maketitle

\section{Introduction}\label{intro}

The Kerr--Newman (KN) family describes stationary electrovacuum black holes in general relativity, characterized by mass, angular momentum, and electric charge~\cite{Carter:1971zc,Ruffini:1971bza}. For a minimally coupled scalar satisfying the usual energy conditions, the no-hair theorems exclude regular scalar hair around asymptotically flat KN black holes~\cite{Bekenstein:1974sf,Bekenstein:1975ts,Bronnikov:1978mx,Sotiriou:2011dz,Herdeiro:2015waa}. Once the scalar is coupled to curvature or matter, the bald solution can instead become marginally unstable and provide the bifurcation point of scalarized solutions. This mechanism was first established for neutron stars in scalar--tensor gravity~\cite{Damour:1993hw} and has since been extended to black holes.

Curvature couplings provide one route to scalarization. In Einstein--scalar--Gauss--Bonnet theory, scalarized branches bifurcate when the linearized equation admits a static scalar cloud~\cite{Kanti:1995vq,Sotiriou:2013qea,Sotiriou:2014pfa,Doneva:2017bvd,Silva:2017uqg,Antoniou:2017acq,Minamitsuji:2018xde}. Rotation can generate a similar spin-induced mechanism for Kerr black holes~\cite{Cunha:2019dwb,Collodel:2019kkx,Dima:2020yac,Hod:2020jjy,Zhang:2020pko,Doneva:2020nbb,Doneva:2020kfv,Herdeiro:2020wei,Berti:2020kgk}. In the Chern--Simons (CS) case, the Pontryagin density contains an overall factor of the spin parameter and is odd under reflection across the equatorial plane, so it couples angular sectors of opposite parity~\cite{Alexander:2009tp,Yunes:2009hc}. The corresponding scalar instability has been studied for Kerr and KN backgrounds~\cite{Myung:2020etf,Zou:2021ybk,Fan:2023jhi}, including the dependence on the azimuthal number and late-time mode mixing in time-domain calculations~\cite{Gao:2018cs}.

Matter couplings give a second route. In Einstein--Maxwell--scalar theory, the Maxwell invariant can destabilize Reissner--Nordstr\"om and KN black holes and can support both positive- and negative-coupling branches when its local sign changes near the polar horizon~\cite{Herdeiro:2018wub,Fernandes:2019rez,Astefanesei:2019pfq,Myung:2018vug,Lai:2022spn,Lai:2022cjs,Guo:2023mda,Hod:2022txa,Hod:2023jdc}. The present model combines this even Maxwell source with the odd CS source. We ask how the two couplings change the axisymmetric response of the bald KN solution, and whether the local sign changes visible in polar cuts coincide with a global instability threshold. We restrict the analysis to linear perturbations and do not construct the nonlinear scalarized solutions.

We use two calculations with different purposes. First, the hyperboloidal R\'acz--T\'oth (RT) evolution places the horizon and future null infinity on a finite computational domain~\cite{Zenginoglu:2011zz,Thuestad:2017ngu,Zenginoglu:2007jw,Racz:2011qu,Harms:2013ib,Harms:2014dqa,Zenginoglu:2010cq} and follows the response to a prescribed Gaussian wave packet. The effective growth or decay rate extracted from a logarithmic-amplitude fit is a finite-time diagnostic of the scalar waveform and can depend on the initial data and on the available evolution time. Second, the stationary zero-mode problem is solved on the full \((r,\theta)\) domain with \(m=0\), regularity at the horizon and on the axes, and decay at infinity. The resulting curves are global axisymmetric marginal-stability curves and are not obtained by interpolating the finite-time waveforms. In particular, \(m=0\) denotes axisymmetry; it does not impose a pure \(\ell=0\) harmonic or reduce the calculation to the local slice \(\theta=0\).

The paper is organized as follows. Section~\ref{sec2} introduces the ECSMS theory, the linearized scalar equation, and the local source structure. Section~\ref{sec3} gives the RT formulation, the Gaussian initial data, the extraction of effective growth and decay rates, and the two-dimensional stationary problem. Section~\ref{sec4results} first presents the time-domain response, then the stationary threshold curves, and finally their comparison at representative points. Section~\ref{sec4} summarizes the results and their limitations. We use geometrized units with \(G=c=1\) and metric signature \((-,+,+,+)\) throughout this work.

\section{ECSMS theory and the linearized scalar equation}\label{sec2}

\subsection{Field equations and local source structure}\label{sec2a}

The Einstein--Chern--Simons--Maxwell--scalar (ECSMS) theory is described by the action
\begin{eqnarray}
S_{\rm ECSMS} = \frac{1}{16\pi} \int \ud^4x \sqrt{-g} \left[ R - 2\nabla_\mu \phi \nabla^\mu \phi + f(\phi) {}^*RR - g(\phi) F_{\mu\nu} F^{\mu\nu} \right], \label{action}
\end{eqnarray}
where $R$ is the Ricci scalar, $\phi$ is the massless scalar field, and $F_{\mu\nu}=\partial_\mu A_\nu-\partial_\nu A_\mu$ is the electromagnetic field strength with $A_\mu$ the $U(1)$ gauge potential. The functions $f(\phi)$ and $g(\phi)$ couple the scalar field to the Pontryagin density and the Maxwell term $F^2\equiv F_{\mu\nu}F^{\mu\nu}$, respectively. The topological Pontryagin density is defined by
\begin{eqnarray}
{}^*RR \equiv \frac{1}{2} \epsilon^{\alpha\beta\gamma\delta} R^\mu{}_{\nu\gamma\delta} R^\nu{}_{\mu\alpha\beta},
\end{eqnarray}
where $\epsilon^{\alpha\beta\gamma\delta}$ is the four-dimensional Levi-Civita tensor and $R^\mu{}_{\nu\alpha\beta}$ denotes the Riemann curvature tensor. We choose the orientation $\epsilon^{tr\theta\varphi}=+1/\sqrt{-g}$. The Pontryagin density is parity odd, so the quadratic coupling used here gives an angular source with a different parity structure from a linear axion coupling.

Variation of the action (\ref{action}) with respect to the metric $g^{\mu\nu}$ yields the Einstein equation
\begin{eqnarray}
G_{\mu\nu} = T^{(\rm CS)}_{\mu\nu} + T^{(\phi)}_{\mu\nu} + T^{(M)}_{\mu\nu}, \label{einsteineq}
\end{eqnarray}
where $G_{\mu\nu} = R_{\mu\nu} - \frac{1}{2}g_{\mu\nu}R$ is the Einstein tensor, and the corresponding stress-energy tensors are given by
\begin{eqnarray}
T^{(\rm CS)}_{\mu\nu} &=& -4\nabla_\sigma f(\phi)\, \epsilon^{\sigma\alpha\beta}{}_{(\mu} \nabla_\beta R_{\nu)\alpha} - 4\nabla_\alpha \nabla_\beta f(\phi)\, {}^*R^\alpha{}_{(\mu\nu)}{}^\beta, \\
T^{(\phi)}_{\mu\nu} &=& 2\nabla_\mu\phi\nabla_\nu\phi - g_{\mu\nu}\nabla_\rho\phi\nabla^\rho\phi, \label{Tphi} \\
T^{(M)}_{\mu\nu} &=& 2g(\phi)\left( F_{\mu\rho}F_\nu{}^\rho - \frac{1}{4}g_{\mu\nu}F^2 \right). \label{TM}
\end{eqnarray}
Varying the action (\ref{action}) with respect to the scalar field $\phi$ yields the modified Klein-Gordon equation
\begin{eqnarray}
\Box\phi + \frac{1}{4}f'(\phi)\,{}^*RR - \frac{1}{4}g'(\phi)F_{\mu\nu}F^{\mu\nu} = 0, \label{kg_orig}
\end{eqnarray}
where $\Box \equiv \nabla_\mu \nabla^\mu$ and a prime denotes differentiation with respect to $\phi$. Varying the action with respect to the electromagnetic potential $A_\mu$ leads to the generalized Maxwell equation
\begin{eqnarray}
\nabla_\mu \left[ g(\phi) F^{\mu\nu} \right] = 0. \label{maxwelleq}
\end{eqnarray}

We study linear scalar perturbations around the bald electrovacuum background $\phi_0 = 0$ and choose the coupling functions as
\begin{eqnarray}
f(\phi) = \gamma \phi^2, \qquad g(\phi) = 1 + \eta \phi^2, \label{couplings}
\end{eqnarray}
which satisfy the relations
\begin{eqnarray}
f(0) = f'(0) = 0, \quad f''(0) = 2\gamma, \qquad g(0) = 1, \quad g'(0) = 0, \quad g''(0) = 2\eta.
\end{eqnarray}
In geometrized units, the length dimensions of the physical quantities are
\begin{eqnarray}
&&[M]=[Q]=[a]=L,\qquad [\phi]=[\eta]=0,\nonumber\\
&&[F^2]=L^{-2},\qquad [{}^*RR]=L^{-4},\qquad [\gamma]=L^2. \label{physical_dimensions}
\end{eqnarray}
The corresponding dimensionless parameters are $a/M$, $Q/M$, $\hat\gamma\equiv\gamma/M^2$, and $\eta$. We set $M=1$ in the numerical calculations below.
For $\eta<0$, the Maxwell kinetic factor $g(\phi)$ remains positive only when $|\phi|<1/\sqrt{|\eta|}$. This condition does not affect the linearization around $\phi=0$, but linear growth alone does not imply a regular nonlinear endpoint.

For $\phi = 0$, both $T^{(\rm CS)}_{\mu\nu}$ and $T^{(\phi)}_{\mu\nu}$ vanish, so the field equations reduce to the Einstein--Maxwell equations and admit the bald KN solution. In Boyer--Lindquist coordinates $(t, r, \theta, \varphi)$, the KN metric is written as
\begin{eqnarray}
d s^2 &=& -\frac{\Delta - a^2\sin^2\theta}{\rho^2}d t^2 - \frac{2a\sin^2\theta(r^2+a^2-\Delta)}{\rho^2}d t d\varphi \nonumber\\
&& + \frac{(r^2+a^2)^2 - \Delta a^2\sin^2\theta}{\rho^2}\sin^2\theta d\varphi^2 + \frac{\rho^2}{\Delta}d r^2 + \rho^2 d\theta^2, \label{kn_metric}
\end{eqnarray}
where
\begin{eqnarray}
\Delta = r^2 - 2Mr + a^2 + Q^2, \qquad \rho^2 = r^2 + a^2\cos^2\theta, \qquad a = \frac{J}{M}.
\end{eqnarray}
The electromagnetic vector potential is given by
\begin{eqnarray}
A = -\frac{Qr}{\rho^2}\left(d t - a\sin^2\theta d\varphi\right).
\end{eqnarray}
The outer horizon $r_+$ and inner horizon $r_-$ are determined by $\Delta = (r-r_+)(r-r_-) = 0$, giving
\begin{eqnarray}
r_\pm = M \pm \sqrt{M^2 - a^2 - Q^2}.\label{kn_horizons}
\end{eqnarray}
The existence of an event horizon requires $a^2 + Q^2 \le M^2$, with the extremal spin given by $a_{\mathrm{e}} = \sqrt{M^2 - Q^2}$.

On the KN background, the Maxwell invariant is
\begin{eqnarray}
F^2 = -\frac{2Q^2\left(r^4 - 6a^2r^2\cos^2\theta + a^4\cos^4\theta\right)}{\left(r^2 + a^2\cos^2\theta\right)^4} \label{maxwell_inv}
\end{eqnarray}
and the Pontryagin density is given by
\begin{eqnarray}
{}^*RR &=& \frac{96 M^2 r a\cos\theta}{\left(r^2 + a^2\cos^2\theta\right)^6}\left[\left(3r^2 - a^2\cos^2\theta\right) - \frac{2Q^2 r}{M}\right] \nonumber\\
&& \times \left[\left(r^2 - 3a^2\cos^2\theta\right) - \frac{Q^2 r}{M}\left(1 - \frac{a^2\cos^2\theta}{r^2}\right)\right]. \label{cs_inv}
\end{eqnarray}
The Maxwell term $F^2$ is even under the equatorial reflection $\theta\to\pi-\theta$, while the factor $\cos\theta$ makes ${}^*RR$ odd. The latter vanishes at $\theta=\pi/2$ and in the nonrotating limit of $a\to0$.
Therefore, reversing $\gamma$ together with $\theta\to\pi-\theta$ leaves the linear spectrum unchanged at fixed $\eta$, $a$, $Q$, and $m$.

For a scalar perturbation $\phi = 0 + \delta\phi$ around the bald KN background, the linearized equation is
\begin{eqnarray}
\left( \Box - \mu_{\rm eff}^2 \right) \delta\phi = 0, \label{linearized_kg}
\end{eqnarray}
where the effective mass squared $\mu_{\rm eff}^2(r,\theta)$ is given by
\begin{eqnarray}
\mu_{\rm eff}^2(r,\theta) = \frac{\eta}{2} F^2 - \frac{\gamma}{2} {}^*RR. \label{eff_mass}
\end{eqnarray}
For a dimensionless comparison of the two source terms, we define
\begin{eqnarray}
\widehat{F}^{2}&=&M^2F^2,\qquad
\widehat{\mathcal P}=M^4{}^*RR,\qquad
\widehat{\mu}_{\rm eff}^{2}=M^2\mu_{\rm eff}^2, \nonumber\\
\widehat{\mu}_{\rm EM}^{2}&=&\frac{\eta}{2}\widehat{F}^{2},\qquad
\widehat{\mu}_{\rm CS}^{2}=-\frac{\hat\gamma}{2}\widehat{\mathcal P},\qquad
\widehat{\mu}_{\rm eff}^{2}=\widehat{\mu}_{\rm EM}^{2}+\widehat{\mu}_{\rm CS}^{2}. \label{dimensionless_mass}
\end{eqnarray}
 At large radius, the two sources behave as
\begin{eqnarray}
&&F^2\sim-\frac{2Q^2}{r^4},\qquad
{}^*RR\sim\frac{288M^2a\cos\theta}{r^7}, \label{source_asymptotics}
\end{eqnarray}
The Maxwell source is even across the equatorial plane, whereas the CS source is odd, and equal values of $\hat\gamma$ and $\eta$ do not give equal source strengths. The numerator of Eq.~(\ref{maxwell_inv}) can change sign near the polar horizon, while $F^2(r,\pi/2)=-2Q^2/r^4$ on the equatorial plane. A region with $\mu_{\rm eff}^2<0$ provides a local indication of a possible tachyonic instability, whereas the existence of a growing global mode must be determined from the full two-dimensional boundary-value problem with the horizon, axis, and asymptotic boundary conditions.

The angular dependence of $\mu_{\rm eff}^2$ couples different multipoles and is the reason that a local polar map cannot replace the global calculation. At fixed azimuthal number $m$, the mass-term part of this coupling can be written as
\begin{eqnarray}
{\cal M}^{(m)}_{\ell\ell'}(r)=\int Y_{\ell m}^*(\theta,\varphi)\,
\mu_{\rm eff}^2(r,\theta)\,Y_{\ell' m}(\theta,\varphi)\ud\Omega. \label{mode_coupling}
\end{eqnarray}
Axisymmetry keeps different $m$ sectors independent, but the off-diagonal entries of Eq.~(\ref{mode_coupling}) mix different values of $\ell$ in a spherical-harmonic basis. The Maxwell term couples harmonics with the same equatorial parity, whereas the CS term couples harmonics with opposite equatorial parities. Here equatorial parity refers to the transformation under $\theta\to\pi-\theta$. These selection rules specify which entries may be nonzero, but not their size. The rotating KN wave operator can introduce additional same-parity mixing, although the minimally coupled equation is separable in a spheroidal-harmonic basis \cite{chandra}. Equation~(\ref{mode_coupling}) describes only the mass-term contribution. Once either source is present, the complete equation is generally non-separable and cannot be represented by a unique one-dimensional radial potential $V_{\rm eff}(r)$. The local mass squared in Eq.~(\ref{dimensionless_mass}) therefore diagnoses where the operator may favour growth, whereas the global criterion is supplied by the boundary-value problem.

In the two-dimensional evolution, these couplings are included directly through the $\theta$-dependent $\mu_{\rm eff}^{2}(r,\theta)$ rather than by constructing the matrix explicitly. An axisymmetric field has the formal expansion $\delta\phi=\sum_{\ell}\phi_\ell(r,t)Y_{\ell0}(\theta)$, so $m=0$ does not imply a pure $\ell=0$ field. We probe this mixing by evolving wave packets proportional to $Y_{\ell_{\rm init}m}$ and projecting the evolved field onto $Y_{\ell m}$; the measured projections contain both source-induced and wave-operator-induced mixing. The stationary calculation used below retains the complete $(r,\theta)$ dependence, while the local plots in this section show only selected northern-hemisphere cuts.
In the following, we discuss the three cases $\hat\gamma=0$, $\hat\gamma=\eta$, and $\hat\gamma\ne\eta$ separately.

\subsection{The \texorpdfstring{$\hat\gamma=0$}{gamma=0} case}\label{sec2b}

\begin{figure}[htb]
\centering
\subfigure[~$\theta=0,~\eta=-10$]{\label{figgamma_zero_negative}\includegraphics[width=2.7in]{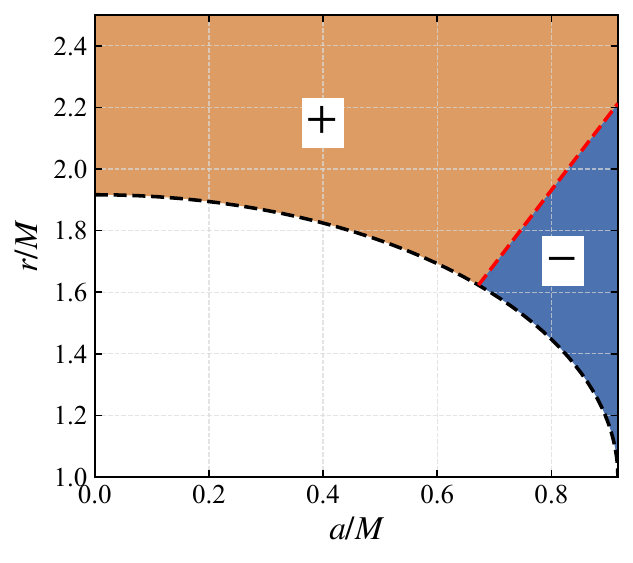}}
\hfill
\subfigure[~$\theta=0,~\eta=1$]{\label{figgamma_zero_positive}\includegraphics[width=2.7in]{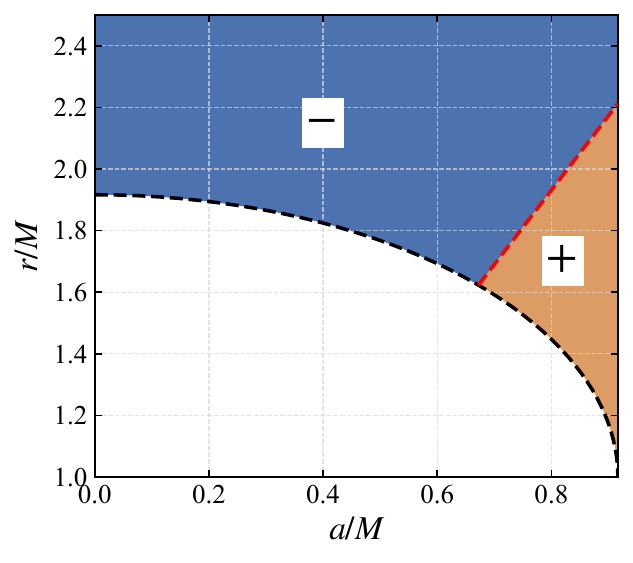}}
\caption{Sign maps of $\widehat\mu_{\rm EM}^{2}$ in the $(a/M,r/M)$ plane, with fixed $M=1$, $Q/M=0.4$, $\hat\gamma=0$, $\theta=0$, and $1\leq r/M\leq2.5$. {Fig.~\ref{figgamma_zero_negative} and Fig.~\ref{figgamma_zero_positive}} correspond to $\eta=-10$ and $\eta=1$, respectively.}\label{figgamma_zero}
\end{figure}

In the pure-Maxwell case, {Eq.~(\ref{dimensionless_mass}) gives $\widehat\mu_{\rm eff}^{2}=\widehat\mu_{\rm EM}^{2}=\eta\widehat F^2/2$.} {Fig.~\ref{figgamma_zero_negative} and Fig.~\ref{figgamma_zero_positive} compare its sign at the northern pole for $\eta=-10$ and $\eta=1$, respectively.} {Blue and orange denote negative and positive values, while the red dashed curves mark the local zeros.} {The dashed black curves show the outer horizon given by Eq.~(\ref{kn_horizons}), and the white regions lie inside the black hole.} From Eq.~(\ref{maxwell_inv}), the exterior zero contour is $r=(1+\sqrt{2})a$, which intersects the outer horizon at $a_{\rm loc}/M=0.6723$ for $Q/M=0.4$. For $\eta<0$, a local region with negative effective mass squared appears only above this spin, while for $\eta>0$ such a region already exists in the nonrotating limit.

Since $a|\cos\theta|/r\leq a/r_+$ throughout the exterior,
$\mu_{\rm eff}^2\geq0$ everywhere for $\hat\gamma=0$, $\eta<0$,
and $a\leq a_{\rm loc}$.
The integral identity in Eq.~(\ref{stationary_integral_identity})
therefore excludes a nontrivial zero mode in this range, so
$a>a_{\rm loc}$ is a necessary condition for the negative-coupling branch.
The global threshold at finite coupling is determined by the stationary
zero-mode analysis in Sec.~\ref{sec3b}.

\FloatBarrier

\subsection{The \texorpdfstring{$\hat\gamma=\eta$}{gamma=eta} case}\label{sec2c}

\begin{figure}[htb]
\centering
\subfigure[~$\theta=0$]{\label{fig_mass_equal_negative_pole}\includegraphics[width=0.29\textwidth]{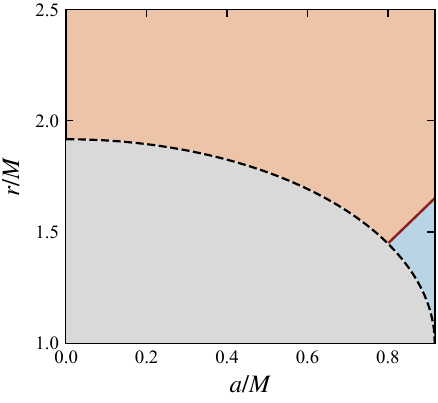}}
\subfigure[~$\theta=\pi/3$]{\label{fig_mass_equal_negative_pi3}\includegraphics[width=0.29\textwidth]{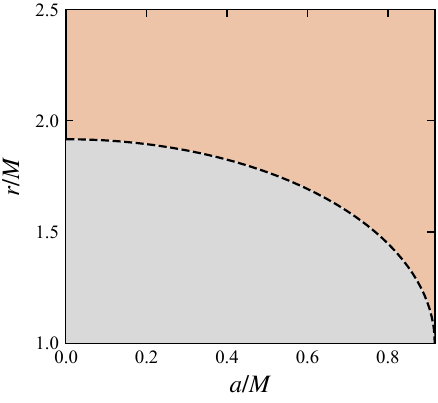}}
\subfigure[~$\theta=0.9\pi/2$]{\label{fig_mass_equal_negative_near_equator}\includegraphics[width=0.29\textwidth]{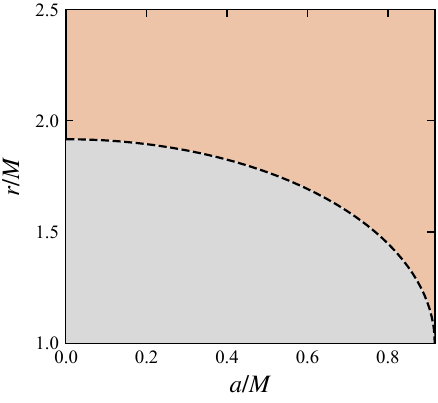}}
\caption{Sign of $\widehat\mu_{\rm eff}^{2}$ for $M=1$, $Q/M=0.4$, $\hat\gamma=\eta=-10$, and $1\leq r/M\leq2.5$. Panels (a), (b), and (c) show the cuts $\theta=0$, $\pi/3$, and $0.9\pi/2$. Blue and orange denote negative and positive regions, while red curves mark local zeros, dashed black curves denote the outer horizon, and gray represents the black-hole interior.}
\label{fig_mass_equal_negative}
\end{figure}

\begin{figure}[htb]
\centering
\subfigure[~$\theta=0$]{\label{fig_mass_equal_positive_pole}\includegraphics[width=0.29\textwidth]{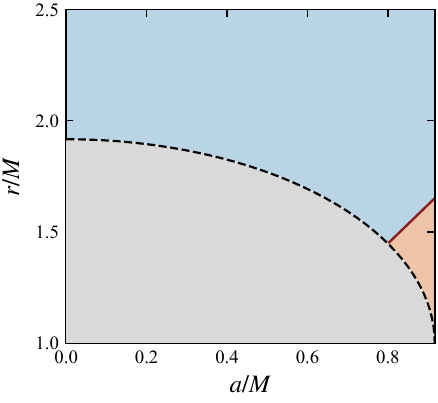}}
\subfigure[~$\theta=\pi/3$]{\label{fig_mass_equal_positive_pi3}\includegraphics[width=0.29\textwidth]{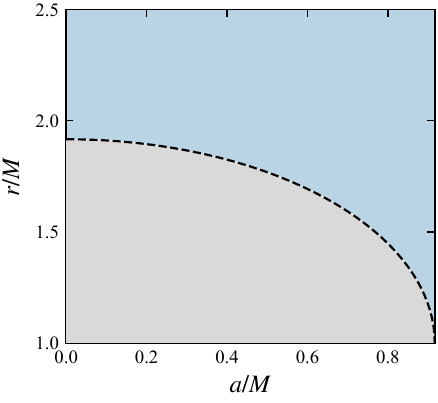}}
\subfigure[~$\theta=0.9\pi/2$]{\label{fig_mass_equal_positive_near_equator}\includegraphics[width=0.29\textwidth]{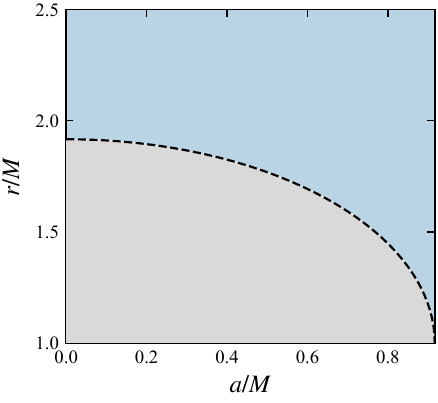}}
\caption{Sign of $\widehat\mu_{\rm eff}^{2}$ for $M=1$, $Q/M=0.4$, $\hat\gamma=\eta=10$, and $1\leq r/M\leq2.5$. Panels (a), (b), and (c) show the cuts $\theta=0$, $\pi/3$, and $0.9\pi/2$. The color and line conventions are the same as in Fig.~\ref{fig_mass_equal_negative}.}\label{fig_mass_equal_positive}
\end{figure}

{For the equal-coupling choice $\hat\gamma=\eta=\lambda$, Eq.~(\ref{dimensionless_mass}) reduces to}
\begin{eqnarray}
\widehat{\mu}_{\rm eff}^{2}=\frac{\lambda}{2}
\left(\widehat F^{2}-\widehat{\mathcal P}\right). \label{equal_mass}
\end{eqnarray}
{For $\lambda\ne0$, the local zeros are determined by the balance $\widehat F^2=\widehat{\mathcal P}$, with the two invariant profiles given in Eqs.~(\ref{maxwell_inv}) and (\ref{cs_inv}).} {Figs.~\ref{fig_mass_equal_negative} and \ref{fig_mass_equal_positive} compare $\lambda=-10$ and $10$ at $M=1$, $Q/M=0.4$, and the three northern-hemisphere cuts $\theta=0$, $\pi/3$, and $0.9\pi/2$.} {As in Fig.~\ref{figgamma_zero}, only points outside the outer horizon in Eq.~(\ref{kn_horizons}) belong to the exterior region considered here.}

For $\lambda=-10$, Fig.~\ref{fig_mass_equal_negative_pole} shows a negative region near the northern polar horizon. {The intersection of the zero contour with the horizon gives the local spin condition for this polar cut:}
\begin{eqnarray}
&&\widehat\mu_{\rm eff}^{2}[r_+(a_{\rm loc}),\theta=0]=0,
\qquad \frac{a_{\rm loc}}{M}\simeq0.8000,\nonumber\\
&&\frac{a_{\rm loc}}{M}<\frac{a}{M}\leq\frac{a_{\rm e}}{M}
=\sqrt{1-\frac{Q^2}{M^2}}\simeq0.9165. \label{local_spin_interval}
\end{eqnarray}
{By contrast, no zero contour occurs at the two sampled off-axis angles in Fig.~\ref{fig_mass_equal_negative_pi3} and Fig.~\ref{fig_mass_equal_negative_near_equator}, where the displayed exterior remains positive.}

\FloatBarrier

{Changing both couplings from $\lambda=-10$ to $10$ reverses the local mass sign according to Eq.~(\ref{equal_mass}), without moving the zero contours.} {Thus, the northern-polar negative region in Fig.~\ref{fig_mass_equal_negative_pole} becomes positive in Fig.~\ref{fig_mass_equal_positive_pole}, while the previously positive off-axis cuts become negative throughout the displayed exterior in Fig.~\ref{fig_mass_equal_positive_pi3} and Fig.~\ref{fig_mass_equal_positive_near_equator}.}

\FloatBarrier

{The northern-polar condition cannot be extended to the whole exterior, since the CS invariant in Eq.~(\ref{cs_inv}) changes sign across the equator while the Maxwell invariant in Eq.~(\ref{maxwell_inv}) does not.} {For example, evaluating Eq.~(\ref{equal_mass}) at the two polar horizon points, with $r_+$ from Eq.~(\ref{kn_horizons}), gives the following values for $\lambda=-10$, $Q/M=0.4$, and $a/M=0.5$:}
\begin{eqnarray}
\widehat\mu_{\rm eff}^{2}(r_+,0)&\simeq&5.25444,\nonumber\\
\widehat\mu_{\rm eff}^{2}(r_+,\pi)&\simeq&-5.12770.\label{polar_mass_example}
\end{eqnarray}
{The southern polar value is negative even though $a/M=0.5$ lies below $a_{\rm loc}/M\simeq0.8$, so the northern-polar condition in Eq.~(\ref{local_spin_interval}) cannot be imposed as a lower bound on the global onset.} {The global equal-coupling thresholds instead follow from the stationary problem in Eq.~(\ref{zero_mode_equation}), with the boundary conditions in Sec.~\ref{sec3b}, and are presented in Fig.~\ref{fig_threshold_v3}.}

\subsection{The \texorpdfstring{$\hat\gamma\ne\eta$}{gamma not equal to eta} case}\label{sec2d}

\begin{figure}[htb]
\centering
\subfigure[~$\theta=0$]{\label{fig_mass_unequal_positive_1_pole}\includegraphics[width=0.29\textwidth]{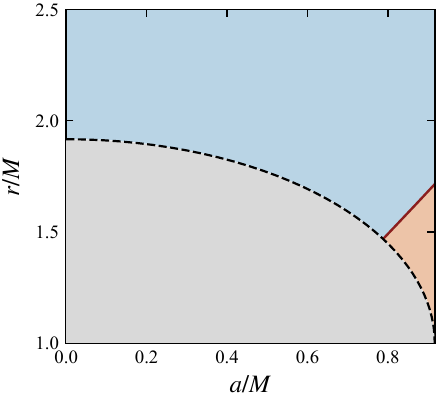}}
\subfigure[~$\theta=\pi/3$]{\label{fig_mass_unequal_positive_1_pi3}\includegraphics[width=0.29\textwidth]{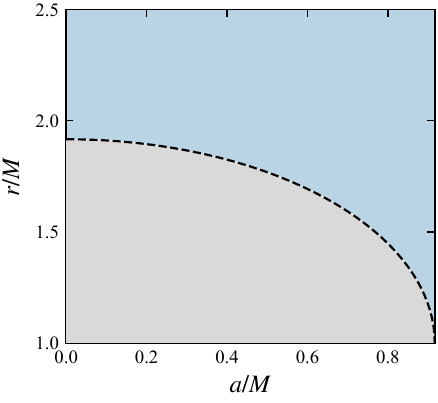}}
\subfigure[~$\theta=0.9\pi/2$]{\label{fig_mass_unequal_positive_1_near_equator}\includegraphics[width=0.29\textwidth]{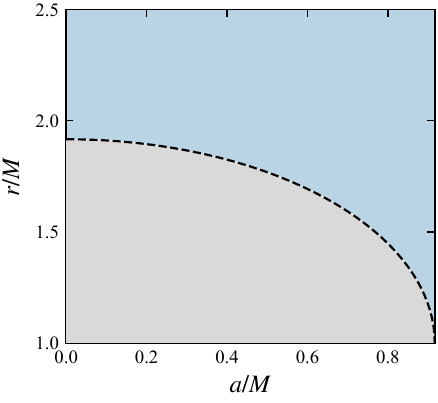}}
\caption{Sign of $\widehat\mu_{\rm eff}^{2}$ for $M=1$, $Q/M=0.4$, $\hat\gamma=1$, $\eta=10$, and $1\leq r/M\leq2.5$.}\label{fig_mass_unequal_positive_1}
\end{figure}

\begin{figure}[htb]
\centering
\subfigure[~$\theta=0$]{\label{fig_mass_unequal_positive_100_pole}\includegraphics[width=0.29\textwidth]{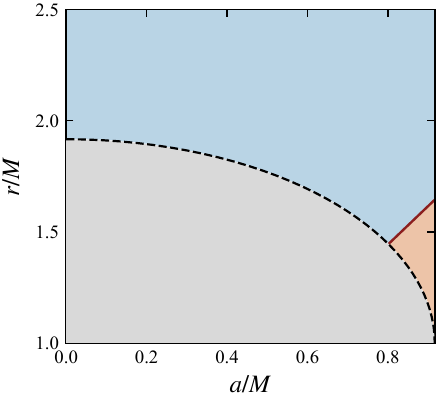}}
\subfigure[~$\theta=\pi/3$]{\label{fig_mass_unequal_positive_100_pi3}\includegraphics[width=0.29\textwidth]{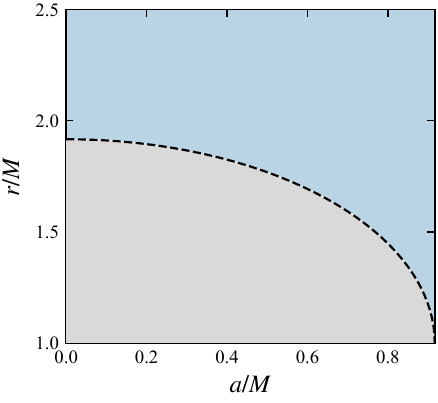}}
\subfigure[~$\theta=0.9\pi/2$]{\label{fig_mass_unequal_positive_100_near_equator}\includegraphics[width=0.29\textwidth]{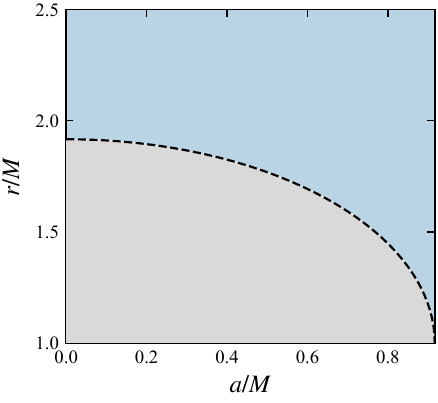}}
\caption{Sign of $\widehat\mu_{\rm eff}^{2}$ for $M=1$, $Q/M=0.4$, $\hat\gamma=100$, $\eta=10$, and $1\leq r/M\leq2.5$.}\label{fig_mass_unequal_positive_100}
\end{figure}

{For independent curvature and electromagnetic couplings, we examine their balance in Eq.~(\ref{dimensionless_mass}) by varying $\hat\gamma$ at fixed positive $\eta$ and then varying negative $\eta$ at fixed $\hat\gamma$.} {All maps below use $M=1$, $Q/M=0.4$, $1\leq r/M\leq2.5$, and the cuts $\theta=0$, $\pi/3$, and $0.9\pi/2$.} {First, Figs.~\ref{fig_mass_unequal_positive_1} and \ref{fig_mass_unequal_positive_100} compare $\hat\gamma=1$ and $100$ at $\eta=10$, so only the CS contribution is varied.} {Increasing $\hat\gamma$ shifts the northern-polar zero contour only slightly, as seen in Fig.~\ref{fig_mass_unequal_positive_1_pole} and Fig.~\ref{fig_mass_unequal_positive_100_pole}, although the magnitude of the CS contribution in Eq.~(\ref{dimensionless_mass}) is much larger.} {At $\theta=\pi/3$ and $0.9\pi/2$, the displayed exterior remains negative for both values of $\hat\gamma$.} The color scale shows only the sign and does not represent this increase in magnitude.

\FloatBarrier

\begin{figure}[htb]
\centering
\subfigure[~$\theta=0$]{\label{fig_mass_eta_m20_pole}\includegraphics[width=0.29\textwidth]{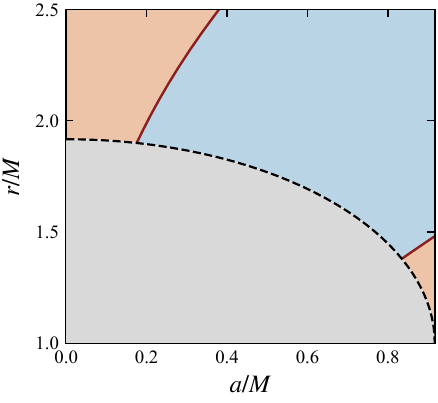}}
\subfigure[~$\theta=\pi/3$]{\label{fig_mass_eta_m20_pi3}\includegraphics[width=0.29\textwidth]{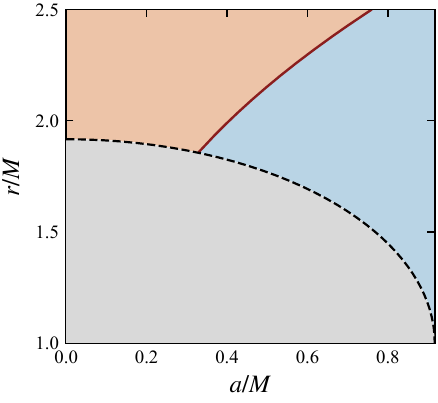}}
\subfigure[~$\theta=0.9\pi/2$]{\label{fig_mass_eta_m20_near_equator}\includegraphics[width=0.29\textwidth]{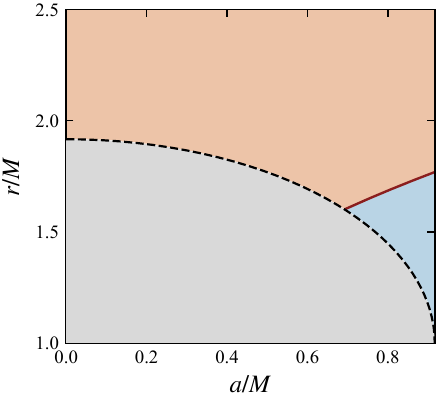}}
\caption{Sign of $\widehat\mu_{\rm eff}^{2}$ for $M=1$, $Q/M=0.4$, $\hat\gamma=1$, $\eta=-20$, and $1\leq r/M\leq2.5$.}\label{fig_mass_eta_m20}
\end{figure}

\begin{figure}[htb]
\centering
\subfigure[~$\theta=0$]{\label{fig_mass_eta_m30_pole}\includegraphics[width=0.29\textwidth]{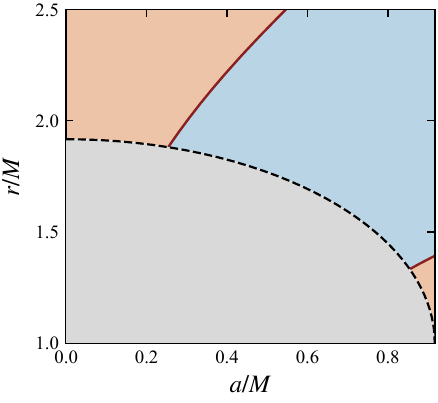}}
\subfigure[~$\theta=\pi/3$]{\label{fig_mass_eta_m30_pi3}\includegraphics[width=0.29\textwidth]{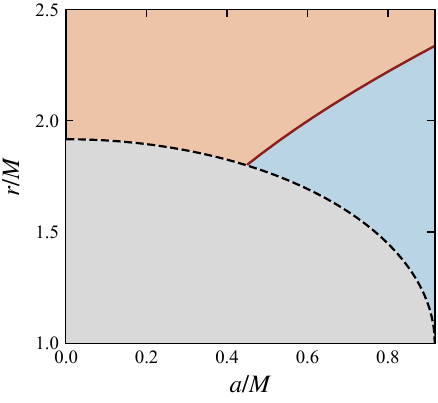}}
\subfigure[~$\theta=0.9\pi/2$]{\label{fig_mass_eta_m30_near_equator}\includegraphics[width=0.29\textwidth]{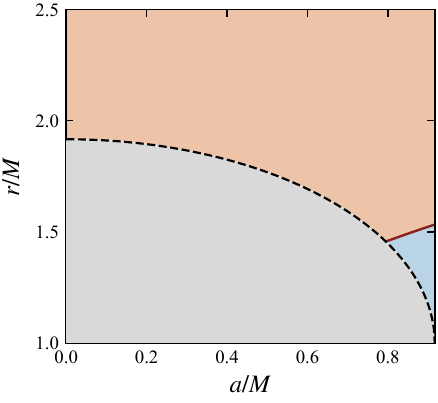}}
\caption{Sign of $\widehat\mu_{\rm eff}^{2}$ for $M=1$, $Q/M=0.4$, $\hat\gamma=1$, $\eta=-30$, and $1\leq r/M\leq2.5$.}\label{fig_mass_eta_m30}
\end{figure}

Next, we fix $\hat\gamma=1$ and take $\eta<0$, for which the Maxwell contribution $\eta\widehat F^2/2$ is positive wherever $F^2<0$ and can oppose a negative CS contribution.For $\eta=-20$ and $-30$, two separated zero-contour segments bound the polar negative region in Fig.~\ref{fig_mass_eta_m20_pole} and Fig.~\ref{fig_mass_eta_m30_pole}, leaving a small positive region near the high-spin polar horizon. For each coupling, moving away from the pole to $\theta=\pi/3$ restricts the negative region to larger spins, as shown in Fig.~\ref{fig_mass_eta_m20_pi3} and Fig.~\ref{fig_mass_eta_m30_pi3}. {Closer to the equatorial plane, Fig.~\ref{fig_mass_eta_m20_near_equator} and Fig.~\ref{fig_mass_eta_m30_near_equator} show only a smaller high-spin negative region.}

\FloatBarrier

\begin{figure}[htb]
\centering
\subfigure[~$\theta=0$]{\label{fig_mass_eta_m70_pole}\includegraphics[width=0.29\textwidth]{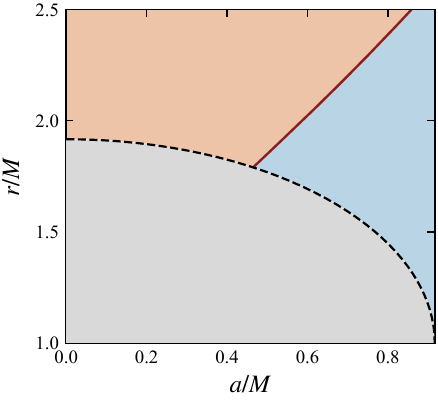}}
\subfigure[~$\theta=\pi/3$]{\label{fig_mass_eta_m70_pi3}\includegraphics[width=0.29\textwidth]{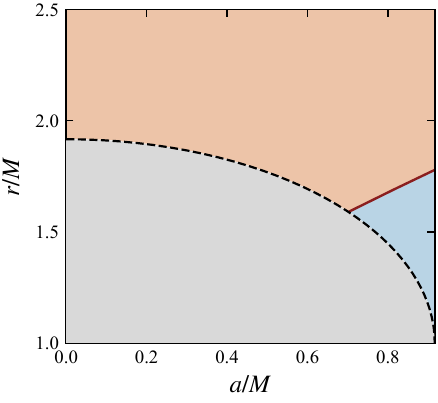}}
\subfigure[~$\theta=0.9\pi/2$]{\label{fig_mass_eta_m70_near_equator}\includegraphics[width=0.29\textwidth]{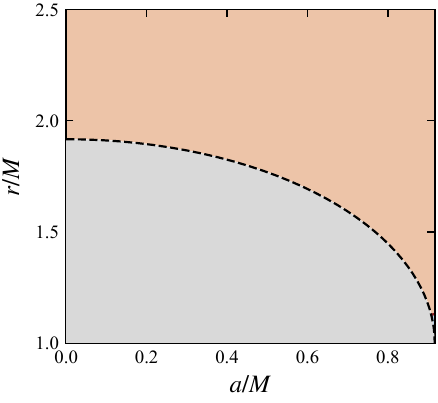}}
\caption{Sign of $\widehat\mu_{\rm eff}^{2}$ for $M=1$, $Q/M=0.4$, $\hat\gamma=1$, $\eta=-70$, and $1\leq r/M\leq2.5$.}\label{fig_mass_eta_m70}
\end{figure}

\begin{figure}[htb]
\centering
\subfigure[~$\theta=0$]{\label{fig_mass_eta_m90_pole}\includegraphics[width=0.29\textwidth]{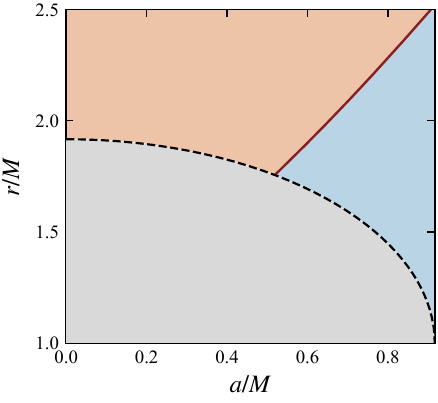}}
\subfigure[~$\theta=\pi/3$]{\label{fig_mass_eta_m90_pi3}\includegraphics[width=0.29\textwidth]{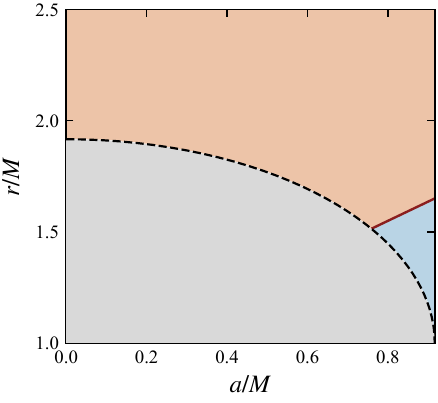}}
\subfigure[~$\theta=0.9\pi/2$]{\label{fig_mass_eta_m90_near_equator}\includegraphics[width=0.29\textwidth]{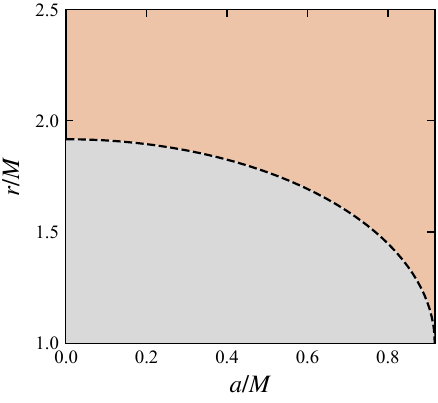}}
\caption{Sign of $\widehat\mu_{\rm eff}^{2}$ for $M=1$, $Q/M=0.4$, $\hat\gamma=1$, $\eta=-90$, and $1\leq r/M\leq2.5$.}\label{fig_mass_eta_m90}
\end{figure}

For the more negative couplings $\eta=-70$ and $-90$, the small positive region near the high-spin polar horizon disappears in Fig.~\ref{fig_mass_eta_m70_pole} and Fig.~\ref{fig_mass_eta_m90_pole}, and the negative region again reaches the near-extremal horizon. At $\theta=\pi/3$, the negative regions instead shrink in Fig.~\ref{fig_mass_eta_m70_pi3} and Fig.~\ref{fig_mass_eta_m90_pi3}, and are confined to a very narrow near-extremal region near the equator in Fig.~\ref{fig_mass_eta_m70_near_equator} and Fig.~\ref{fig_mass_eta_m90_near_equator}. {This angular contrast follows from Eq.~(\ref{dimensionless_mass}) together with the Maxwell invariant in Eq.~(\ref{maxwell_inv}): increasing $|\eta|$ strengthens a positive Maxwell contribution where $F^2<0$, but a negative one where $F^2>0$ near the polar horizon.}

\FloatBarrier

These maps identify local negative-mass regions, whose presence alone does not determine the time-domain response. For example, both the $\eta=-20$ and $-30$ maps contain negative regions, whereas the corresponding waveforms at $a/M=0.9$ in Fig.~\ref{fig_wave_q04} grow and decay, respectively.
We next describe the time-domain evolution and stationary zero-mode
methods used to determine the dynamical response and global thresholds.

\section{\texorpdfstring{Numerical method}{Numerical method}}\label{sec3}

\subsection{Hyperboloidal evolution and numerical consistency checks}\label{sec3a}

For the time-domain evolution, the ingoing Kerr--Schild-type coordinates are related to the Boyer--Lindquist coordinates by
\begin{eqnarray}
\ud\widetilde{t} = \ud t+\frac{2Mr-Q^2}{\Delta}\ud r, \quad
\ud\widetilde{\varphi} =\ud\varphi+\frac{a}{\Delta}\ud r. \label{ingoing_coord}
\end{eqnarray}
The first relation gives the horizon-regular time $\widetilde t=v-r$. The axial symmetry of the KN background allows each azimuthal sector to evolve independently through the decomposition
\begin{eqnarray}
\delta\phi(\widetilde t,r,\theta,\widetilde\varphi)
&=&\frac{\psi(\widetilde t,r,\theta)}{r}e^{im\widetilde\varphi}. \label{fourier_dec}
\end{eqnarray}
The numerical calculations use $M=1$. In the RT formulas below, $r$, $\widetilde t$, $T$, and $h$ are measured in units of $M$, whereas $R$ is dimensionless. The RT transformation is
\begin{eqnarray}
\Omega &=& \frac{1-R^2}{2}, \quad
r = \frac{R}{\Omega}=\frac{2R}{1-R^2}, \nonumber\\
\widetilde t &=& T+h(R), \quad
h(R) =\frac{1+R^2}{2\Omega}-4M\ln(2\Omega). \label{rt_coord}
\end{eqnarray}
The event horizon and future null infinity are located at
\begin{eqnarray}
R_+ &=& \frac{r_+}{\sqrt{1+r_+^2}+1},\qquad R_{\mathscr I^+}=1, \label{rt_domain}
\end{eqnarray}
respectively. Thus, the finite interval $R\in[R_+,1]$ covers the complete exterior region and includes future null infinity at its outer endpoint. No artificial timelike boundary is introduced.

After division by the coefficient of $\partial_T^2\psi$, the compactified equation takes the form
\begin{eqnarray}
0 &=& \partial_T^2\psi+c_{TR}\partial_T\partial_R\psi
+c_{RR}\partial_R^2\psi+c_{\theta\theta}\partial_\theta^2\psi \nonumber\\
&& +c_T\partial_T\psi+c_R\partial_R\psi
+c_\theta\partial_\theta\psi+c_0\psi. \label{rt_pde}
\end{eqnarray}
The coefficients depend on the background geometry and the azimuthal number $m$, while the scalar couplings enter $c_0$ through the effective mass in Eq.~(\ref{eff_mass}). They are written in terms of the auxiliary quantities
\begin{align*}
r&=\frac{2R}{1-R^2},\\
\mathcal J&=\frac{\ud R}{\ud r}=\left(\frac{\ud r}{\ud R}\right)^{-1},\\
H&=\frac{\ud h}{\ud r}=\mathcal J\frac{\ud h}{\ud R},\\
\mathcal D&=-2(r^2+a^2)(1+H)+\Delta(1+H)^2+a^2\sin^2\theta,\\
\mathcal K&=2Mr-Q^2-\Delta H.
\end{align*}
After substituting the transformed inverse metric, the coefficients in Eq.~(\ref{rt_pde}) are
\begin{align*}
 c_{TR}&=\frac{2\mathcal K\mathcal J}{\mathcal D},\qquad
 c_{RR}=\frac{\Delta\mathcal J^2}{\mathcal D},\qquad
 c_{\theta\theta}=\frac{1}{\mathcal D},\\
 c_T&=\frac{\partial_r\mathcal K-2\mathcal K/r-2\mathrm{i}maH}{\mathcal D},\qquad
 c_R=\frac{\partial_r(\Delta\mathcal J)-2\Delta\mathcal J/r+2\mathrm{i}ma\mathcal J}{\mathcal D},\\
 c_\theta&=\frac{\cot\theta}{\mathcal D},\\
 c_0&=\frac{1}{\mathcal D}\left[
 \frac{2\Delta}{r^2}-\frac{\partial_r\Delta}{r}-m^2\csc^2\theta
 -\rho^2\mu_{\rm eff}^2-\frac{2\mathrm{i}ma}{r}\right].
\end{align*}
For $m=0$, the coefficients and real initial data remain real. For $m\ne0$, the terms proportional to $\mathrm{i}m$ couple the real and imaginary parts, which are evolved together in the state vector below.
Defining
\begin{eqnarray}
\chi = \partial_R\psi, \quad
\Pi = \partial_T\psi, \label{aux_vars}
\end{eqnarray}
the first-order-in-time system becomes
\begin{eqnarray}
\partial_T\psi &=& \Pi, \nonumber\\
\partial_T\chi &=& \partial_R\Pi, \label{pde_system}\\
\partial_T\Pi &=& -\big(c_{TR}\partial_R\Pi+c_{RR}\partial_R\chi
+c_{\theta\theta}\partial_\theta^2\psi \nonumber\\
&& \hspace{1.0cm}+c_T\Pi+c_R\chi+c_\theta\partial_\theta\psi+c_0\psi\big). \nonumber
\end{eqnarray}
Writing each of $\psi$, $\chi$, and $\Pi$ in terms of its real and imaginary parts, the state vector is given by
\begin{eqnarray}
&&\bm u=\left(\psi_{\mathrm R},\chi_{\mathrm R},\Pi_{\mathrm R},
\psi_{\mathrm I},\chi_{\mathrm I},\Pi_{\mathrm I}\right)^{\mathsf T}. \nonumber
\end{eqnarray}
The initial data are a localized Gaussian wave packet with angular factor $Y_{\ell_{\rm init}m}$:
\begin{eqnarray}
\psi(0,R,\theta)&=&\mathcal A_0Y_{\ell_{\rm init}m}(\theta,0)
\exp\!\left[-\frac{(R-R_c)^2}{2\sigma^2}\right],\nonumber\\
\chi(0,R,\theta)&=&\partial_R\psi(0,R,\theta),\qquad
\Pi(0,R,\theta)=0,\label{gaussian_initial_data}
\end{eqnarray}
Here $R_c=(R_++1)/2$, $\sigma=1/\sqrt{200}$, and $\mathcal A_0=1$. The normalization is immaterial for the linear equation. The background conserves $m$, but it does not conserve $\ell$: the rotating wave operator and the angular dependence of $\mu_{\rm eff}^2$ mix the harmonics during evolution. Equation~(\ref{pde_system}) is integrated using the fourth-order Runge--Kutta method with $C_{\mathrm{CFL}}=0.30$. The representative checks of Figs.~\ref{fig_gamma0_branches}, \ref{fig_threshold_v3}, and \ref{fig_scan_comparison} use $\ell_{\rm init}=m=0$, $(N_R,N_\theta)=(97,32)$, $T_{\max}/M=850$, and $\theta_{\rm obs}=\pi/4$, with fits over $450\leq T/M\leq800$. The weak-growth evolutions at $\eta=-24$ and $-82.75$ use $(N_R,N_\theta)=(351,64)$ with time steps $\Delta T/M=0.004504$ and $0.004491$, respectively, and the waveforms are sampled at intervals of approximately $M$.

The radial grid includes both endpoints $R=R_+$ and $R=1$, while the $N_\theta$ angular points are cell-centered and exclude the ghost cells. The corresponding grid spacings are given by
\begin{eqnarray}
&&\Delta R=\frac{1-R_+}{N_R-1},\qquad
\Delta\theta=\frac{\pi}{N_\theta}. \nonumber
\end{eqnarray}
The angular nodes are
\begin{eqnarray}
\theta_j &=& \left(j-\frac12\right)\Delta\theta,\qquad
j=1,\ldots,N_\theta, \label{angular_grid}
\end{eqnarray}
Two ghost cells are placed beyond each pole. Seven-point radial and five-point angular stencils give sixth- and fourth-order accuracy in the interior, respectively, with one-sided radial stencils at the endpoints.
Smoothness at the axes requires $\psi,\chi,\Pi=O(\sin^{|m|}\theta)$. The corresponding parity extension for the ghost cells is
\begin{eqnarray}
\bm u(T,R,-\theta) &=& (-1)^m \bm u(T,R,\theta), \nonumber\\
\bm u(T,R,\pi+\theta) &=& (-1)^m \bm u(T,R,\pi-\theta). \label{axis_bc}
\end{eqnarray}
The radial characteristic speeds are given by
\begin{eqnarray}
\lambda_\pm &=& \frac{c_{TR}\pm\sqrt{c_{TR}^2-4c_{RR}}}{2}. \label{characteristic_speeds}
\end{eqnarray}
At both radial endpoints, $c_{RR}=0$ and the nonzero characteristic speed points out of the domain, so no incoming radial data are imposed.

For the near-marginal evolutions, the real and imaginary fields at $\theta_{\rm obs}$ are linearly interpolated between neighbouring angular nodes at $R=1$. A fit to the logarithmic scalar amplitude over a finite time window gives the effective growth or decay rate $\Gamma$:
\begin{eqnarray}
\mathcal A(T)&\equiv&|\psi(T,1,\theta_{\mathrm{obs}})|,\qquad
\ln\mathcal A(T)=c+\Gamma T. \label{fit_formula}
\end{eqnarray}
For $m\ne0$, $\psi$ is complex but $\mathcal A$ is real. A positive (negative) $\Gamma$ therefore indicates growth (decay) of the extracted amplitude over the selected interval. This finite-time rate need not coincide with the imaginary part of a converged quasinormal-mode frequency, particularly near threshold where transients or late-time tails contribute to the signal. Since $T$ and $M$ have the same dimension, Table~\ref{tab:growth_rates_q04} reports the dimensionless combinations $M\Gamma$ and $M\delta\Gamma$. The raw amplitude and an envelope with nominal bin width $(T_{\max}-T_{\min})/15$ are fitted over
\begin{eqnarray}
&&100\leq T/M\leq250\qquad(\hbox{ringdown}),\nonumber\\
&&250\leq T/M\leq450\qquad(\hbox{transition}),\nonumber\\
&&450\leq T/M\leq800\qquad(\hbox{late time}). \nonumber
\end{eqnarray}
The envelope keeps the largest discretely sampled amplitude in each bin, and a sample at $T_{\max}$ is retained as an endpoint bin. The fit uncertainty is defined as
\begin{eqnarray}
\delta\Gamma &=& 2\sigma_\Gamma+\Delta_{\mathrm{win}}, \label{fit_error}
\end{eqnarray}
Here $\sigma_\Gamma$ is the least-squares error from ordinary linear regression \cite{Press:2007}, while $\Delta_{\mathrm{win}}$ is the largest change after shifting either window edge by ten percent of its width \cite{Bhagwat:2017tkm,Giesler:2019uxc,Witek:2018dmd}. The quoted $\delta\Gamma$ accounts only for regression and fitting-window variation, without including spatial or time-step errors or establishing that a single mode has been isolated. Growth or decay is identified over the fitted interval only when the raw and envelope fits agree in sign and $|\Gamma|>\delta\Gamma$. A negative slope that approaches zero is recorded as a tail rather than as evidence for a strictly damped eigenmode.

For the axisymmetric evolutions, projection of $\psi(T,1,\theta)$ onto $Y_{\ell0}$ at $\mathscr I^+$ gives the multipolar coefficients $C_\ell(T)$. The normalized multipolar power fraction is measured by
\begin{eqnarray}
P_\ell(T)&=&\frac{\langle|C_\ell|^2\rangle_{\Delta T}}
{\displaystyle\sum_{j=0}^{4}\langle|C_j|^2\rangle_{\Delta T}},
\label{angular_power_fraction}
\end{eqnarray}
The average is taken over a moving interval of width $\Delta T\simeq11M$. The index $\ell$ labels a component of the evolved field, whereas $\ell_{\rm init}$ in Eq.~(\ref{gaussian_initial_data}) specifies only the initial angular multipole. The denominator contains the five retained harmonics $0\leq\ell\leq4$, so $P_\ell$ is a normalized multipolar power fraction at infinity and is not an energy fraction of the full perturbation.

For the decaying evolution $(a/M,Q/M,\hat\gamma,\eta)=(0.9,0.4,1,-30)$, spatial convergence is checked at three resolutions, $(N_R,N_\theta)=(176,32)$, $(351,64)$, and $(701,128)$, over $100\leq T/M\leq250$. Writing $\Gamma_{N_R,N_\theta}$ for the fitted rate at each resolution, the observed convergence order is estimated as\cite{Baumgarte:2010,Sperhake:2014wpa,Roache:1998}
\begin{eqnarray}
p &=& \log_2\left|
\frac{\Gamma_{176,32}-\Gamma_{351,64}}
{\Gamma_{351,64}-\Gamma_{701,128}}
\right|\simeq3.29. \label{convergence_order}
\end{eqnarray}
The three grids contain $175$, $350$, and $700$ radial intervals. The sampling times make a small contribution to $p\simeq3.29$, while interpolation to common times gives $p\simeq4.22$. This convergence test concerns the early damped signal and does not establish the discretization error of the two weak late-time growth rates.

\subsection{Stationary zero modes and threshold couplings}\label{sec3b}

A stationary axisymmetric zero mode is independent of $t$ and $\varphi$, with $\omega=m=0$, and can be written as $\delta\phi=\Phi(r,x)$,
where $x=\cos\theta$ and $\rho^2=r^2+a^2x^2$. The full $(r,x)$ dependence is retained, so no truncation to a single spherical harmonic is imposed. Substituting $\delta\phi=\Phi(r,x)$ into Eq.~(\ref{linearized_kg}) gives
\begin{eqnarray}
\partial_r\!\left(\Delta\partial_r\Phi\right)
+\partial_x\!\left[(1-x^2)\partial_x\Phi\right]
-\rho^2\mu_{\rm eff}^2\Phi=0.\label{zero_mode_equation}
\end{eqnarray}
The boundary conditions are regularity of $\Phi$ at the outer horizon $r=r_+$ and both symmetry axes $x=\pm1$, together with decay to zero at spatial infinity. For real $\Phi$, integrating Eq.~(\ref{zero_mode_equation}) by parts gives
{
\begin{eqnarray}
\int_{r_+}^{\infty}\!\ud r\int_{-1}^{1}\!\ud x\,
\left[\Delta(\partial_r\Phi)^2+(1-x^2)(\partial_x\Phi)^2
+\rho^2\mu_{\rm eff}^2\Phi^2\right]=0.
\label{stationary_integral_identity}
\end{eqnarray}
}
{The boundary terms vanish under these conditions, and the nonnegative gradient terms in Eq.~(\ref{stationary_integral_identity}) exclude a nontrivial zero mode if $\mu_{\rm eff}^2\geq0$ throughout the exterior.}

{For the equal-coupling case $\hat\gamma=\eta=\lambda$ with $M=1$, substituting Eq.~(\ref{eff_mass}) into Eq.~(\ref{zero_mode_equation}) gives}
\begin{eqnarray}
\partial_r\!\left(\Delta\partial_r\Phi\right)
+\partial_x\!\left[(1-x^2)\partial_x\Phi\right]
&=&\lambda\,\frac{\rho^{2}}{2}
\left(F^{2}-{}^{*}RR\right)\Phi .\label{equal_zero_mode_equation}
\end{eqnarray}
Here the invariants are expressed in units of $M$, corresponding to $\widehat F^2$ and $\widehat{\mathcal P}$ in Eq.~(\ref{dimensionless_mass}). The eigenvalues $\lambda$ give the equal-coupling thresholds.

The stationary boundary-value problem is solved by Chebyshev pseudospectral collocation in the compact radial coordinate $s=1/r$ and the angular coordinate $x$, over $s\in[0,1/r_+]$ and $x\in[-1,1]$. The Chebyshev--Lobatto grids include all endpoints, and the asymptotic values $\Phi=0$ at $s=0$ are eliminated from the unknowns. At the horizon and both axes, including the corner points, Eq.~(\ref{zero_mode_equation}) is imposed in its regular limiting form. For independent couplings with fixed $\hat\gamma$ and $M=1$, the Maxwell term in Eq.~(\ref{eff_mass}) is linear in $\eta$, so the discretized equation takes the generalized eigenvalue form

\begin{eqnarray}
A\boldsymbol{\Phi}=\eta B\boldsymbol{\Phi},\qquad
B=\frac{1}{2}\operatorname{diag}\!\left(\rho^2F^2\right).
\label{stationary_matrix_problem}
\end{eqnarray}
Here $\boldsymbol{\Phi}$ contains the unknown field values at the collocation points, while $A$ discretizes the derivative terms in Eq.~(\ref{zero_mode_equation}) plus the fixed CS term $\gamma\rho^2{}^*RR/2$. A real eigenvalue is accepted as a threshold only if its residual is small, its eigenfunction has no radial or angular nodes, and the spectrum of $A-\eta B$ has no positive eigenvalue above the numerical tolerance. This spectral check selects a zero of the principal spatial mode, rather than an excited zero mode on an already unstable background. The critical couplings are obtained with $N_s=22$ and $N_x=21$ and checked with $N_s=26$ and $N_x=25$, using higher resolutions for the low-spin negative branches and spin-dependent mergers. Branch continuation at fixed $a/M$ and $Q/M$ is based on normalized overlaps between neighboring eigenfunctions with the $dr\,dx$ measure.

{For $a/M=0.9$ and $Q/M=0.4$, solving Eq.~(\ref{stationary_matrix_problem}) gives the following nodeless real roots for the pure-Maxwell case $\hat\gamma=0$ and the mixed-coupling case $\hat\gamma=1$:}
{
\begin{eqnarray}
\left.(\eta_0^{(1)},\eta_0^{(2)})\right|_{\hat\gamma=0}
&=&(20.5987,-102.8439),\nonumber\\
\left.(\eta_0^{(1)},\eta_0^{(2)})\right|_{\hat\gamma=1}
&=&(-24.0552,-82.2081).\label{stationary_roots}
\end{eqnarray}
}
The roots in Eq.~(\ref{stationary_roots}) change by less than $2\times10^{-6}$ between the two grids, below the displayed precision. The stationary roots define the marginal couplings, while the RT evolutions of Sec.~\ref{sec3a} test the associated growth or decay at selected points.

\section{\texorpdfstring{Results}{Results}}\label{sec4results}

The time-domain response, mode mixing, and stationary thresholds obtained with the methods of Sec.~\ref{sec3} are presented below.

\subsection{Time-domain response}\label{sec4m0}\label{sec4mneq0}

\begin{figure}[htb]
\centering
\includegraphics[width=6.2in]{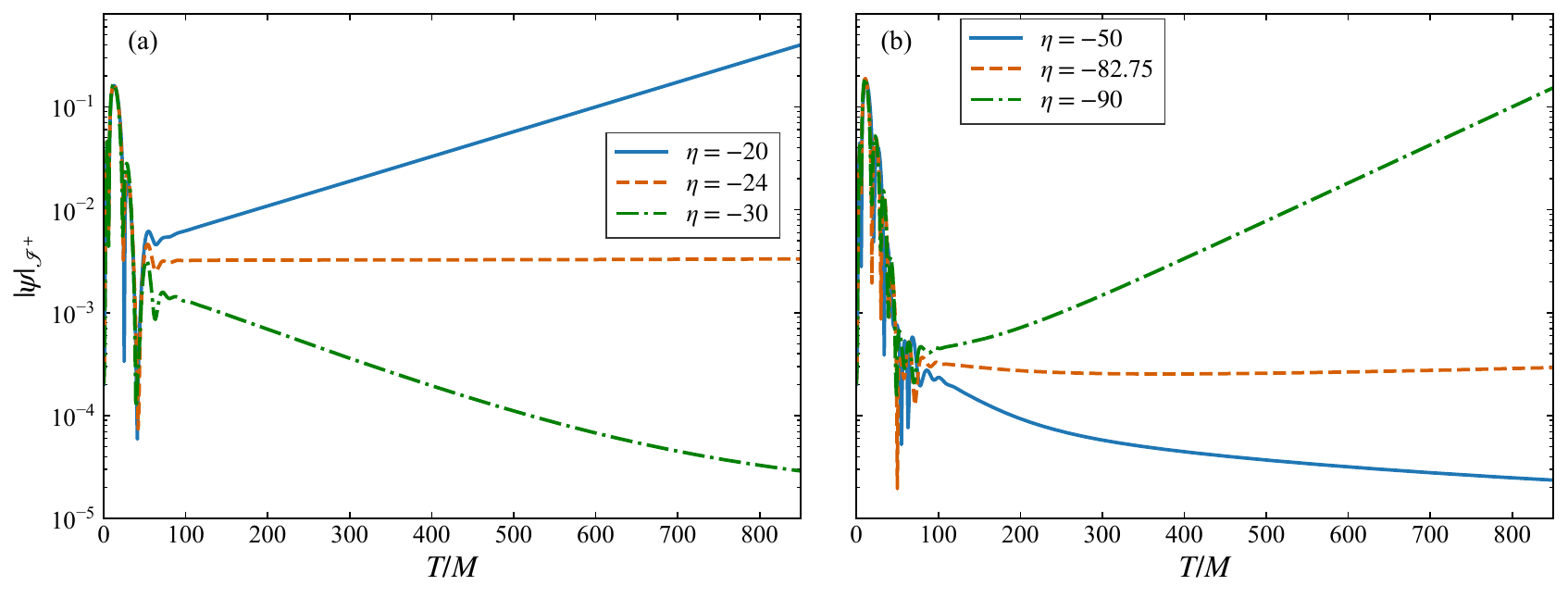}
\caption{Axisymmetric amplitudes at $\mathscr I^+$ with fixed $M=1$, $a/M=0.9$, $Q/M=0.4$, $\hat\gamma=1$, $\ell_{\rm init}=m=0$, and $\theta_{\rm obs}=\pi/4$. Panels (a) and (b) correspond to $\eta=-20,-24,-30$ and $\eta=-50,-82.75,-90$, respectively. The amplitudes at $\eta=-20,-90$ grow, while those at $\eta=-30,-50$ decay and develop tails.}\label{fig_wave_q04}
\end{figure}

For $m=0$, Fig.~\ref{fig_wave_q04} shows the response at $a/M=0.9$, $Q/M=0.4$, $\hat\gamma=1$, and $\ell_{\rm init}=0$, using $(N_R,N_\theta)=(351,64)$. The signals at $\eta=-20$ and $-90$ grow, while those at $\eta=-30$ and $-50$ decay and develop late-time tails. The near-threshold signals at $\eta=-24$ and $-82.75$ have much smaller late-time slopes, with the latter changing from early decay to weak growth. The raw and envelope fits in Table~\ref{tab:growth_rates_q04} have the same sign over $450\leq T/M\leq800$. These responses reflect the competition between the fixed CS source and the Maxwell contribution to the two-dimensional effective mass, although the finite-time fitted rates alone do not determine the threshold.

\begin{table}[htb]
\caption{Finite-time amplitude fits for the signals in Fig.~\ref{fig_wave_q04}, with fixed $M=1$, $a/M=0.9$, $Q/M=0.4$, $\hat\gamma=1$, $\ell_{\rm init}=m=0$, and $\theta_{\rm obs}=\pi/4$. The raw columns use the discretely sampled amplitudes directly, while the envelope column uses the bin maxima defined in Sec.~\ref{sec3a}. The quoted uncertainties follow Eq.~(\ref{fit_error}) and describe the regression and fitting-window variation.}\label{tab:growth_rates_q04}
\centering
\footnotesize
\begin{tabular}{cccc}
\hline\hline
$\eta$ & \makecell{$M\Gamma_{100\text{--}250}$\\raw} & \makecell{$M\Gamma_{450\text{--}800}$\\raw} & \makecell{$M\Gamma_{450\text{--}800}$\\envelope} \\
\hline
$-20$    & $+5.516\times10^{-3}$ & $(5.5521\pm0.0004)\times10^{-3}$ & $(5.5523\pm0.0007)\times10^{-3}$ \\
$-24$    & $+4.705\times10^{-5}$ & $(5.327\pm0.128)\times10^{-5}$ & $(5.386\pm0.238)\times10^{-5}$ \\
$-30$    & $-6.492\times10^{-3}$ & $(-4.272\pm0.202)\times10^{-3}$ & $(-4.371\pm0.410)\times10^{-3}$ \\
$-50$    & $-7.973\times10^{-3}$ & $(-1.373\pm0.049)\times10^{-3}$ & $(-1.396\pm0.105)\times10^{-3}$ \\
$-82.75$ & $-1.397\times10^{-3}$ & $(3.349\pm0.179)\times10^{-4}$ & $(3.474\pm0.295)\times10^{-4}$ \\
$-90$    & $+5.426\times10^{-3}$ & $(8.508\pm0.005)\times10^{-3}$ & $(8.511\pm0.009)\times10^{-3}$ \\
\hline\hline
\end{tabular}
\normalsize
\end{table}

\FloatBarrier

\begin{figure}[htb]
\centering
\subfigure[$a/M=0.9$ and $Q/M=0.1$]{\label{fig_wave_q01}\includegraphics[width=2.7in]{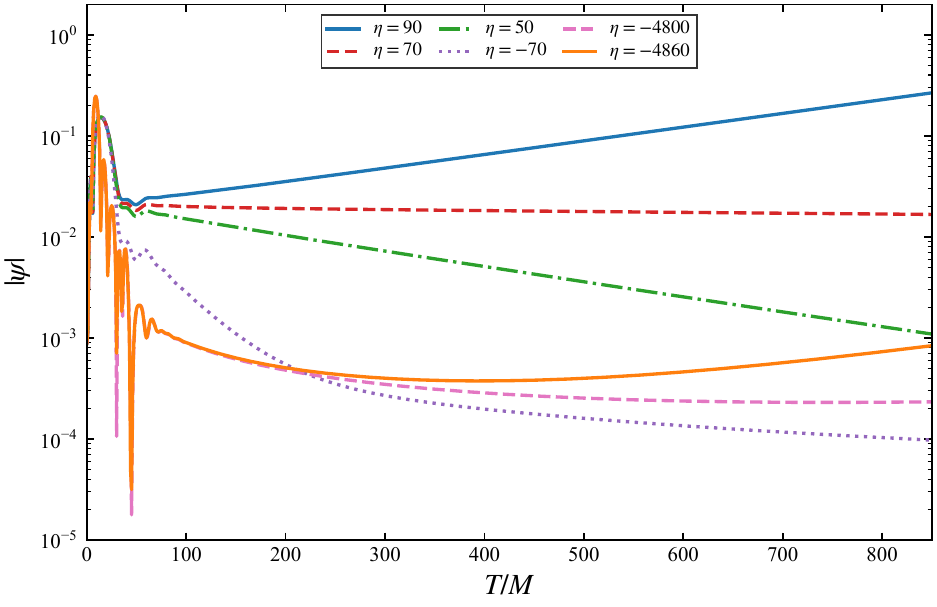}}
\hfill
\subfigure[$a/M=0.5$ and $Q/M=0.4$]{\label{fig_wave_a05}\includegraphics[width=2.7in]{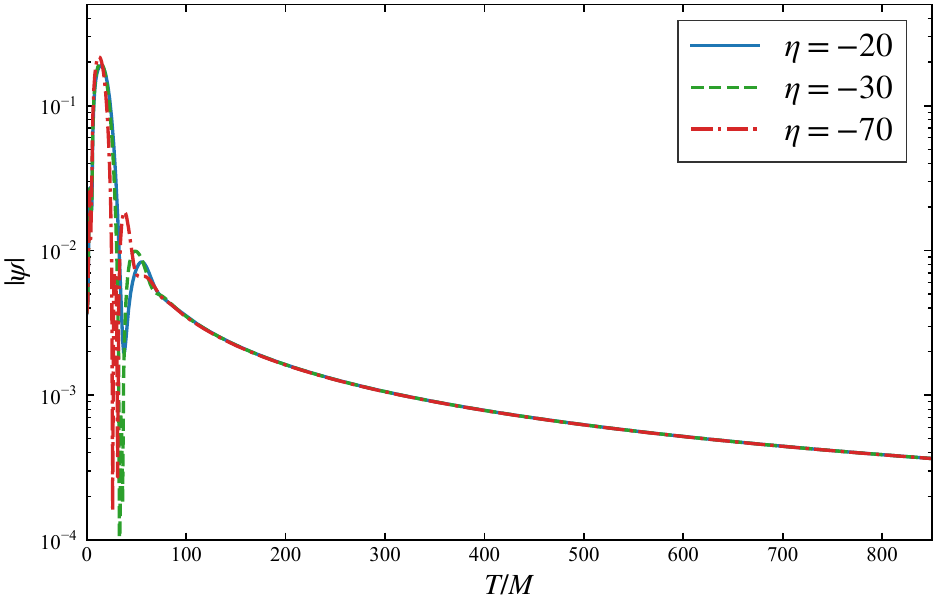}}
\caption{Axisymmetric amplitudes at $\mathscr I^+$ with fixed $M=1$, $\hat\gamma=1$, $\ell_{\rm init}=m=0$, and $\theta_{\rm obs}=\pi/4$. Panel (a) shows $\eta=90,70,50,-70,-4800,-4860$ for $a/M=0.9$ and $Q/M=0.1$, while panel (b) shows $\eta=-20,-30,-70$ for $a/M=0.5$ and $Q/M=0.4$. The smaller-spin cases decay over the calculated time interval.}\label{fig_wave_pair}
\end{figure}

The charge and spin scans in Fig.~\ref{fig_wave_pair} illustrate why the local effective mass alone does not determine global stability.For the negative-coupling samples at $Q/M=0.1$ and $a/M=0.9$ in Fig.~\ref{fig_wave_q01}, growth occurs at $\eta=-4860$, whereas the fitted slope at $\eta=-4800$ changes sign with the fitting window. The positive-coupling sample $\eta=90$ also grows. The factor $Q^2$ in Eq.~(\ref{maxwell_inv}) explains the weaker Maxwell source, although changing $Q/M$ also changes the distance from extremality and therefore does not define a charge-scaling law by itself. At $a/M=0.5$ and $Q/M=0.4$, all three sampled scalar perturbations decay through $T/M=850$. The spin changes both the source profiles and the wave operator, so these time-domain samples are used as dynamical observations rather than as global linear instability thresholds.

\begin{figure}[htb]
\centering
\includegraphics[width=3.65in]{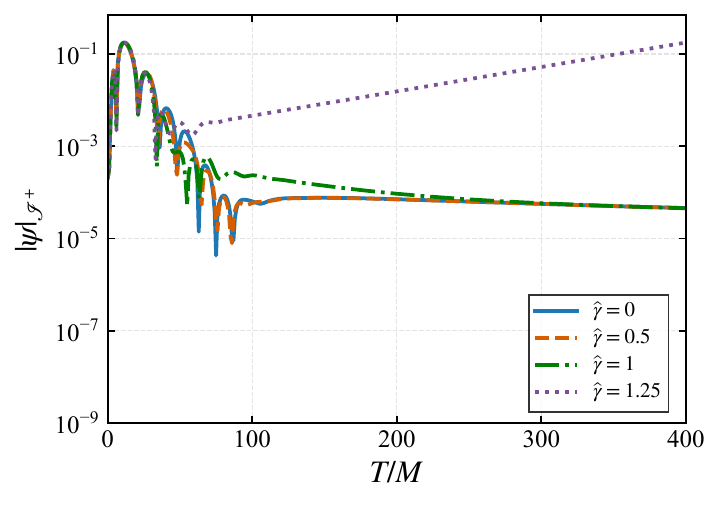}
\caption{Axisymmetric amplitudes at $\mathscr I^+$ with fixed $M=1$, $a/M=0.9$, $Q/M=0.4$, $\eta=-50$, $\ell_{\rm init}=m=0$, and $\theta_{\rm obs}=\pi/4$, for $\hat\gamma=0,0.5,1,1.25$. The amplitude grows at $\hat\gamma=1.25$ and decreases at the three smaller sampled couplings.}\label{fig_coupling_scan}
\end{figure}

At fixed $a/M=0.9$, $Q/M=0.4$, and $\eta=-50$, increasing $\hat\gamma$ also changes the response. In Fig.~\ref{fig_coupling_scan}, the signals for $\hat\gamma=0,0.5,1$ decrease through $T/M=400$, whereas the signal for $\hat\gamma=1.25$ grows. Here only the CS contribution to the effective mass is varied, which isolates its role in the competition with the Maxwell term.

\begin{figure}[htb]
\centering
\includegraphics[width=3.65in]{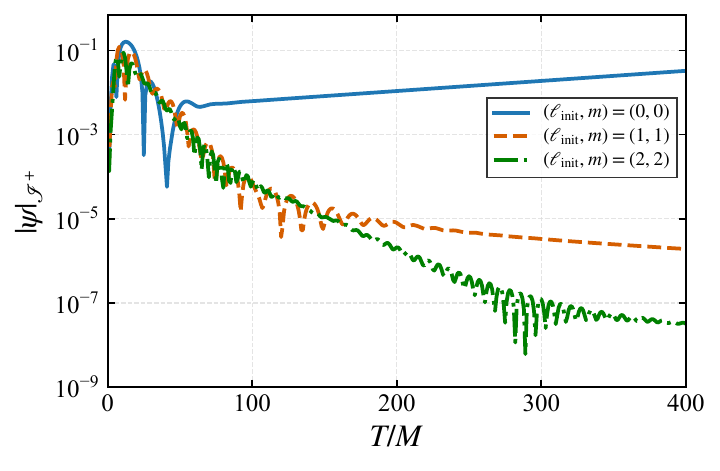}
\caption{Amplitudes at $\mathscr I^+$ with fixed $M=1$, $a/M=0.9$, $Q/M=0.4$, $(\hat\gamma,\eta)=(1,-20)$, and $\theta_{\rm obs}=\pi/4$. The three curves correspond to the lowest allowed initial multipoles $(\ell_{\rm init},m)=(0,0),(1,1),(2,2)$, respectively. The $m=0$ amplitude grows, while the $m=1,2$ amplitudes decay through $T/M=400$.}\label{fig_m_modes}
\end{figure}

Fig.~\ref{fig_m_modes} shows the responses for $m=0,1,2$ with the lowest allowed initial multipole $\ell_{\rm init}=|m|$, corresponding to $(\ell_{\rm init},m)=(0,0),(1,1),(2,2)$. At $a/M=0.9$, $Q/M=0.4$, and $(\hat\gamma,\eta)=(1,-20)$, the $m=0$ signal grows while the $m=1$ and $m=2$ signals decrease through $T/M=400$. Extending the $m=1,2$ runs at $\eta=-20,-50,-90$ to $T/M=850$ gives decaying tails in all six cases. The decay observed for these initial data over a finite time does not establish stability of the $m=1,2$ sectors, which requires separate stationary or spectral threshold calculations.

\FloatBarrier

\subsection{Mode mixing and initial-data dependence}\label{sec4checks}

\begin{figure}[htb]
\centering
\subfigure[$\ell_{\rm init}=0$]{\label{ERR1}\includegraphics[width=2.8in]{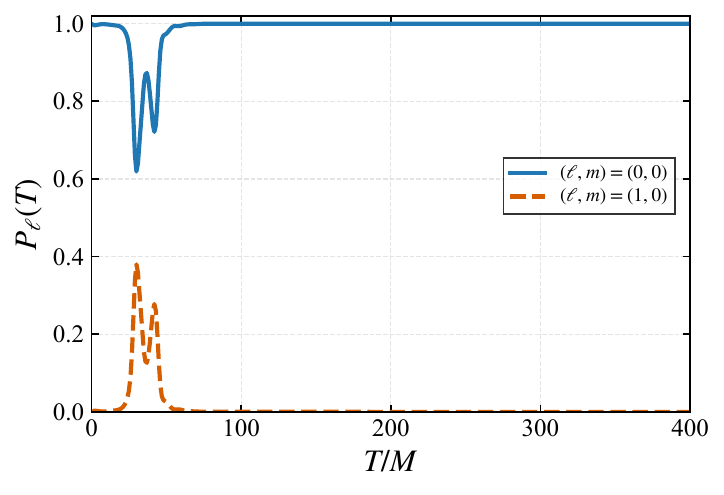}}
\hfill
\subfigure[$\ell_{\rm init}=1$]{\label{ERR2}\includegraphics[width=2.8in]{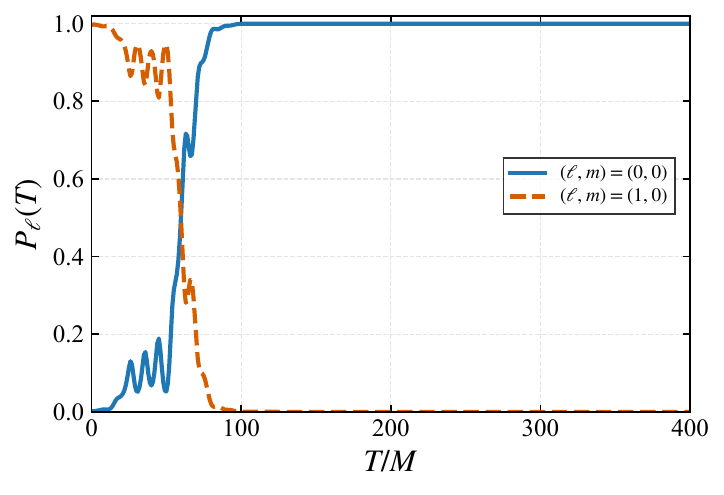}}

\subfigure[$\ell_{\rm init}=2$]{\label{ERR3}\includegraphics[width=2.8in]{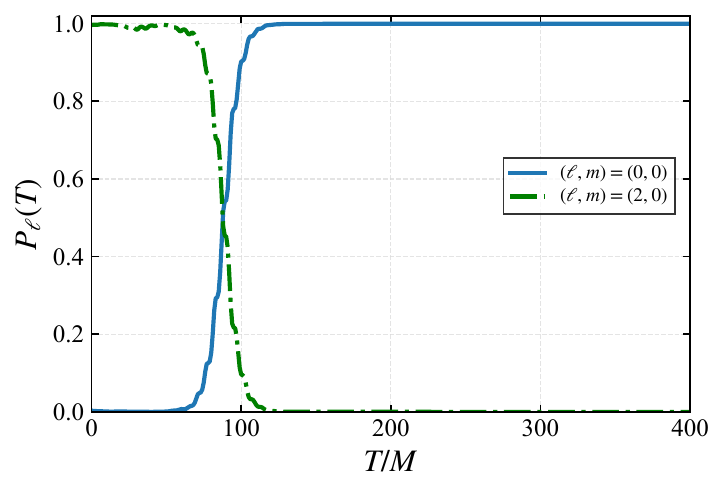}}
\hfill
\subfigure[$\ell_{\rm init}=3$]{\label{ERR4}\includegraphics[width=2.8in]{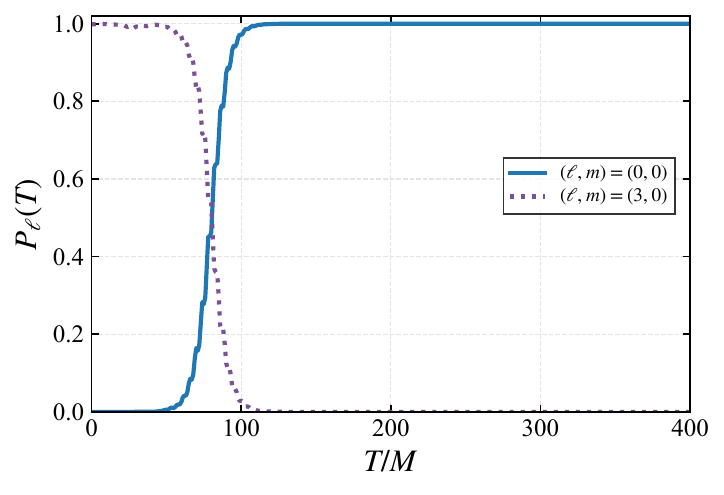}}
\caption{Normalized multipolar power fractions at $\mathscr I^+$ for $M=1$, $a/M=0.9$, $Q/M=0.4$, $(\hat\gamma,\eta)=(1,-20)$, and $m=0$, with $\ell_{\rm init}=0,1,2,3$ in panels (a)--(d). The normalization retains $0\leq\ell\leq4$.}\label{figERR}
\end{figure}

The dependence on the initial angular profile is shown in Fig.~\ref{figERR} for wave packets proportional to $Y_{\ell_{\rm init}0}$ with $\ell_{\rm init}=0,1,2,3$ at $a/M=0.9$, $Q/M=0.4$, and $(\hat\gamma,\eta)=(1,-20)$, using $(N_R,N_\theta)=(176,32)$. The two largest fractions defined in Eq.~(\ref{angular_power_fraction}) are displayed, with the normalization retaining $0\leq\ell\leq4$. All four initial multipoles yield
\begin{eqnarray}
&&M\Gamma\simeq5.55\times10^{-3},
\qquad 250\leq T/M\leq380,\nonumber\\
&&P_0(T/M=400)>0.99,
\qquad \ell_{\rm init}=0,1,2,3. \label{late_time_mode_summary}
\end{eqnarray}
This implies that the late-time growth is dominated by the monopole component with nearly the same rate for all four initial multipoles, although their transient responses differ. The rotating wave operator and the Maxwell source preserve equatorial parity, whereas the odd CS source couples even and odd harmonics within the same $m$ sector through Eq.~(\ref{mode_coupling}). Consequently, an initially odd wave packet can also excite the $\ell=0$ component. This monopole dominance refers to the evolved field, while the stationary zero-mode calculation retains the full angular dependence. Higher harmonics and the convergence of the fractions with angular resolution have not been tested.

Close to a threshold, the finite-time rate also depends on the initial radial profile. For $\eta=-24$ and $-82.75$, the late-time slopes reported in Table~\ref{tab:growth_rates_q04} for the original Gaussian initial data remain positive under the tested changes of fitting window, resolution, time step, and observation angle. Changing the radial centre or packet width delays the visible growth, so the modified packets initially give decaying fits and show growth only in longer evolutions, with different fitted rates. The critical coupling is therefore determined from the stationary equation, while the time-domain evolution tests the associated dynamical response.
\FloatBarrier

\subsection{Critical couplings and stationary zero modes}\label{sec4threshold}

\begin{figure}[!htbp]
\centering
\includegraphics[width=4in]{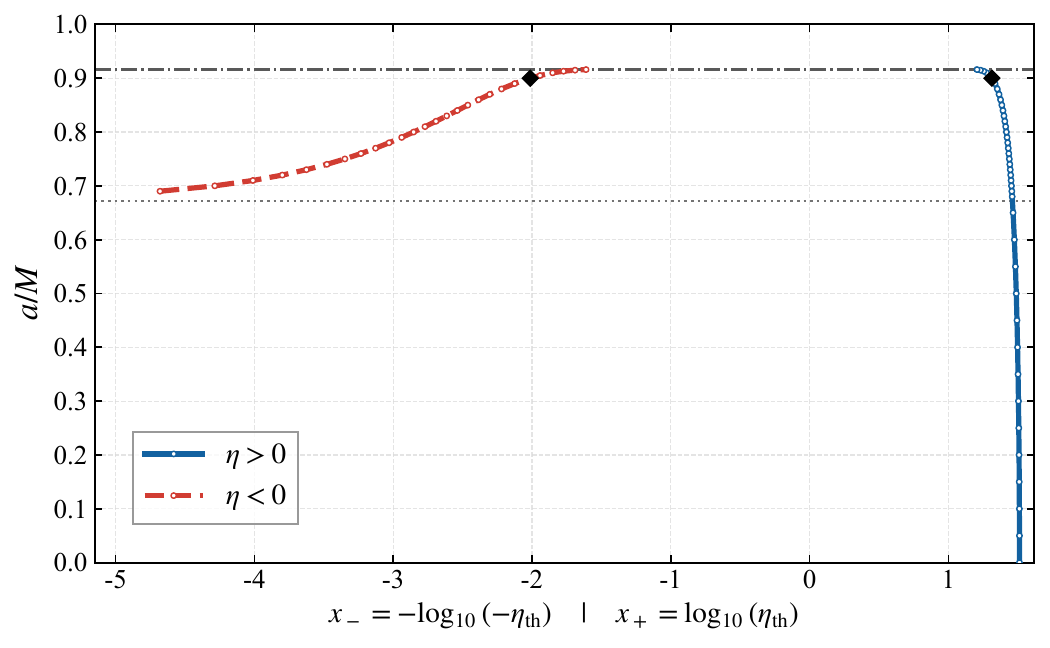}
\caption{Global axisymmetric $m=0$ zero-mode branches with fixed $M=1$, $Q/M=0.4$, and $\hat\gamma=0$. Blue solid and red dashed curves join the computed positive- and negative-coupling samples, respectively. Black diamonds mark $\eta=20.5987$ and $-102.8439$ at $a/M=0.9$. The dotted and dash-dotted horizontal lines mark the necessary spin bound $a_{\rm loc}/M=0.6723$ and the extremal spin $a_{\rm e}/M=0.9165$. {The resolved negative branch covers} $0.69\leq a/M\leq0.916$ and remains above $a_{\rm loc}$, while no extrapolation to either limiting spin is shown.}\label{fig_gamma0_branches}
\end{figure}

For the pure-Maxwell case $\hat\gamma=0$, the representative thresholds are given in Eq.~(\ref{stationary_roots}).
The corresponding global axisymmetric branches are shown in Fig.~\ref{fig_gamma0_branches}, using the stationary eigenvalues with the convergence and principal-root checks described in Sec.~\ref{sec3b}. The negative- and positive-coupling branches are displayed in the separate logarithmic coordinates $x_-=-\log_{10}(-\eta_{\rm th})$ and $x_+=\log_{10}(\eta_{\rm th})$, respectively, where $\eta_{\rm th}$ denotes the threshold coupling. The plotted points retain the full two-dimensional angular dependence, and the connecting lines join adjacent samples in these coordinates. At $a/M=0.9$, the two marked points appear at $\eta=20.5987$ and $-102.8439$, which are the same roots quoted above. For $\eta<0$, the necessary sign-change condition of Fig.~\ref{figgamma_zero} requires $a_{\rm loc}/M<a/M<a_{\rm e}/M$, where $a_{\rm loc}/M=0.6723$ and $a_{\rm e}/M\simeq0.9165$.  All retained negative-branch points satisfy this bound. The resolved negative branch covers $0.69\leq a/M\leq0.916$, and no continuation toward $a_{\rm loc}$ or the extremal limit is inferred from the connecting lines. The positive-coupling branch also includes the nonrotating limit, with $\eta=32.5549$ at $a=0$. The two branches are consistent with the pure-Maxwell results in Refs.~\cite{Lai:2022spn,Lai:2022cjs}.

\begin{figure}[H]
\centering
\includegraphics[width=2.7in]{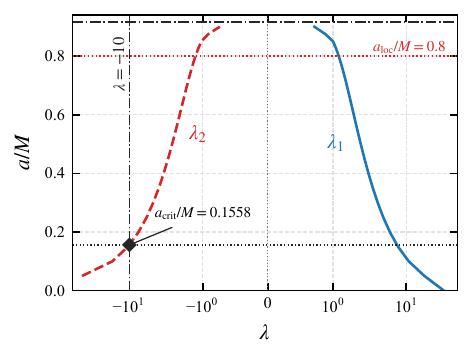}
\caption{Global axisymmetric $m=0$ zero-mode thresholds for $\hat\gamma=\eta=\lambda$, with fixed $M=1$ and $Q/M=0.4$. The black diamond marks the $\lambda=-10$ onset. Dotted and dash-dotted lines indicate the northern-polar local sign change and extremal spin, respectively.}
\label{fig_threshold_v3}
\end{figure}

For $\hat\gamma=\eta=\lambda$, Fig.~\ref{fig_threshold_v3} shows the global thresholds obtained from Eq.~(\ref{equal_zero_mode_equation}), with the blue solid and red dashed curves corresponding to $\lambda_1>0$ and $\lambda_2<0$, respectively.
Both threshold magnitudes decrease with spin, reaching $\lambda_1=0.7144$ and $\lambda_2=-0.7370$ at $a/M=0.9$, so rotation allows zero modes to appear at weaker couplings.
The branches are not symmetric under $\lambda\to-\lambda$: equatorial reflection reverses the odd CS source in Eq.~(\ref{cs_inv}) but leaves the even Maxwell source in Eq.~(\ref{maxwell_inv}) unchanged, whereas reversing $\lambda$ reverses both contributions.

For a fixed negative $\lambda$, its intersection with $\lambda_2(a/M)$ determines the onset spin.
The vertical line $\lambda=-10$ meets this branch at the black diamond, giving
\begin{eqnarray}
\left.\frac{a_{\rm crit}}{M}\right|_{\lambda=-10}&\simeq&0.1558,
\label{equal_lambda_onset}
\end{eqnarray}
This onset lies well below the northern-polar local sign-change spin $a_{\rm loc}/M\simeq0.8$ in Eq.~(\ref{local_spin_interval}), indicated by the red dotted line.
The odd CS source produces a negative-mass region in the southern exterior at lower spins, as illustrated in Eq.~(\ref{polar_mass_example}), so the northern-polar value is not a lower bound on the global onset.
The dash-dotted line marks the KN extremal spin $a_{\rm e}/M\simeq0.9165$ from Eq.~(\ref{kn_horizons}), while the threshold curves are computed only up to $a/M=0.9$.

\begin{figure}[H]
\centering
\subfigure[~$\hat\gamma=1$]{\label{fig_scan_gamma1}\includegraphics[width=2in]{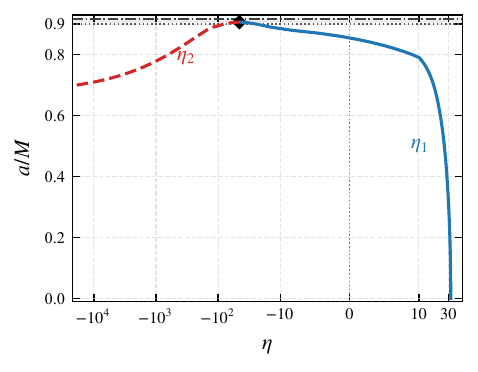}}
\hfill
\subfigure[~$\hat\gamma=5$]{\label{fig_scan_gamma5}\includegraphics[width=2in]{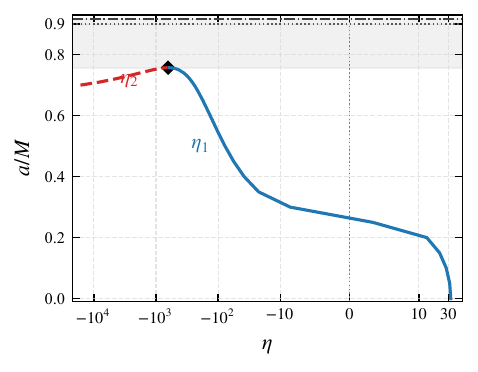}}
\hfill
\subfigure[~$a/M=0.9$]{\label{fig_scan_stationary}\includegraphics[width=2in]{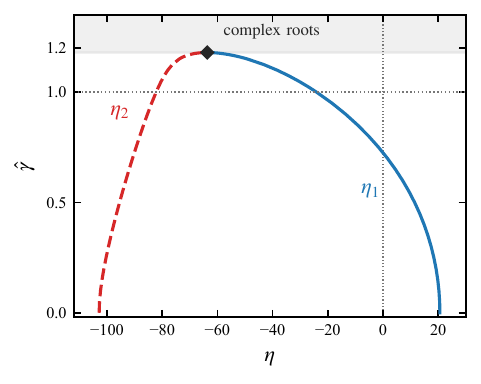}}
\caption{Global axisymmetric $m=0$ stationary zero-mode branches with fixed $M=1$ and $Q/M=0.4$. Blue solid and red dashed curves show $\eta_1$ and $\eta_2$. Black diamonds and gray bands mark mergers and complex coupling roots, respectively.}\label{fig_scan_comparison}
\end{figure}

For unequal CS and Maxwell couplings, Fig.~\ref{fig_scan_gamma1} and Fig.~\ref{fig_scan_gamma5} show the two real threshold branches $\eta_1$ and $\eta_2$ as functions of spin at fixed $\hat\gamma=1$ and $5$, respectively. The pure-Maxwell reference branches are shown in Fig.~\ref{fig_gamma0_branches}.
At $\hat\gamma=1$, $\eta_1$ starts at the positive Maxwell root at $a=0$ and becomes negative as spin increases, while $\eta_2$ enters the displayed range from more negative values.
At $a/M=0.9$, the roots are those given in Eq.~(\ref{stationary_roots}).
The RT checks in Table~\ref{time_domain_threshold_checks} show decay for sampled couplings between the two roots and growth for those outside them.
The two branches approach each other and meet at the black diamonds, with

\begin{eqnarray}
\left.\left(\frac{a_{\rm coll}}{M},\eta_{\rm coll}\right)\right|_{\hat\gamma=1}
&\simeq&(0.90665,-45.76),\nonumber\\
\left.\left(\frac{a_{\rm coll}}{M},\eta_{\rm coll}\right)\right|_{\hat\gamma=5}
&\simeq&(0.75714,-638.1).
\label{spin_branch_mergers}
\end{eqnarray}

A stronger CS coupling therefore brings the merger to a smaller spin.
The merger spins agree at the quoted precision on the three checked grids, up to $(N_s,N_x)=(30,29)$.
The lower endpoints reflect the plotted range $-22000\leq\eta\leq50$ and the available samples, so they do not determine minimum onset spins.

At fixed $a/M=0.9$, Fig.~\ref{fig_scan_stationary} instead shows how the thresholds change with $\hat\gamma$.
The branch $\eta_1$ crosses $\eta=0$ at the pure-CS threshold $\hat\gamma=0.7257$, obtained from Eq.~(\ref{zero_mode_equation}) with $\eta=0$.
Beyond this point, both roots are negative and approach each other as $\hat\gamma$ increases, merging at the black diamond near
\begin{eqnarray}
(\hat\gamma_{\rm coll},\eta_{\rm coll})\simeq(1.18,-63.69),
\label{branch_merger}
\end{eqnarray}
Beyond the mergers, the tracked roots form complex-conjugate pairs, shown by the gray bands.
These are coupling eigenvalues of Eq.~(\ref{stationary_matrix_problem}), rather than mode frequencies, so their complex continuation does not determine stability at real coupling.

The stationary thresholds are compared with independent RT evolutions at representative points. All checks in Tables~\ref{equal_lambda_time_domain_checks} and \ref{time_domain_threshold_checks} use $M=1$ and $Q/M=0.4$. The Gaussian initial data are given by Eq.~(\ref{gaussian_initial_data}) with $\ell_{\rm init}=m=0$, and the fitted rates follow Eqs.~(\ref{fit_formula}) and (\ref{fit_error}).The evolutions use $(N_R,N_\theta)=(97,32)$ and run to $T_{\max}/M=850$, with waveforms extracted at $\theta_{\rm obs}=\pi/4$. The raw amplitude and its envelope are fitted over $450\leq T/M\leq800$. The columns $M\Gamma_{\rm raw}$ and $M\Gamma_{\rm env}$ give the rates obtained from the sampled amplitudes and the bin maxima, respectively, using the diagnostics defined in Sec.~\ref{sec3a}. All listed samples satisfy the sign criterion of Sec.~\ref{sec3a}: the raw and envelope fits agree in sign and both have $|\Gamma|>\delta\Gamma$. For the equal-coupling case $\lambda=-10$, the fitted rates at $a/M=0.15$ and $0.16$ are listed in Table~\ref{equal_lambda_time_domain_checks}.
\begin{table}[H]
\caption{Finite-time growth and decay rates for the equal-coupling onset checks with fixed $M=1$, $Q/M=0.4$, and $\hat\gamma=\eta=\lambda=-10$.}\label{equal_lambda_time_domain_checks}
\centering
\footnotesize
\begin{tabular}{crr}
\hline\hline
$a/M$ & $M\Gamma_{\rm raw}$ & $M\Gamma_{\rm env}$ \\
\hline
$0.15$ & $-2.0753\times10^{-3}$ & $-2.0803\times10^{-3}$ \\
$0.16$ & $+6.0316\times10^{-3}$ & $+6.0326\times10^{-3}$ \\
\hline\hline
\end{tabular}
\normalsize
\end{table}

The opposite signs in Table~\ref{equal_lambda_time_domain_checks} are consistent with the stationary onset in Eq.~(\ref{equal_lambda_onset}).For the $\hat\gamma=5$ check at $a/M=0.75$, Eq.~(\ref{stationary_matrix_problem}) gives the stationary roots $\eta_1=-428.1270$ and $\eta_2=-1135.2966$. The fixed-CS rates are summarized in Table~\ref{time_domain_threshold_checks}. These growth and decay signs support the correspondence between the stationary thresholds and the sampled dynamics.For $\eta=-50$ at $\hat\gamma=1$, Tables~\ref{tab:growth_rates_q04} and \ref{time_domain_threshold_checks} give different decay rates on grids $(N_R,N_\theta)=(351,64)$ and $(97,32)$, respectively. Both rates indicate decay, although these tail slopes are not converged mode frequencies. Near marginality, changes in the radial profile delay the visible growth, so the finite-time $\Gamma$ in Eq.~(\ref{fit_formula}) does not determine an exact stationary onset.

\begin{table}[H]
\caption{Finite-time growth and decay rates for the fixed-CS branch checks with fixed $M=1$ and $Q/M=0.4$.}\label{time_domain_threshold_checks}
\centering
\footnotesize
\begin{tabular}{cccrr}
\hline\hline
$\hat\gamma$ & $a/M$ & $\eta$ & $M\Gamma_{\rm raw}$ & $M\Gamma_{\rm env}$ \\
\hline
\multirow{3}{*}{$0$} & \multirow{3}{*}{$0.9$} & $-130$ & $+3.1805\times10^{-2}$ & $+3.1805\times10^{-2}$ \\
 & & $0$ & $-3.4446\times10^{-4}$ & $-3.5877\times10^{-4}$ \\
 & & $25$ & $+2.1987\times10^{-2}$ & $+2.1987\times10^{-2}$ \\
\hline
\multirow{3}{*}{$1$} & \multirow{3}{*}{$0.9$} & $-20$ & $+5.5582\times10^{-3}$ & $+5.5584\times10^{-3}$ \\
 & & $-50$ & $-5.1583\times10^{-4}$ & $-5.3730\times10^{-4}$ \\
 & & $-90$ & $+8.5139\times10^{-3}$ & $+8.5157\times10^{-3}$ \\
\hline
\multirow{3}{*}{$5$} & \multirow{3}{*}{$0.75$} & $-400$ & $+4.6374\times10^{-3}$ & $+4.6977\times10^{-3}$ \\
 & & $-700$ & $-1.0363\times10^{-3}$ & $-1.0681\times10^{-3}$ \\
 & & $-1300$ & $+6.3310\times10^{-3}$ & $+6.7324\times10^{-3}$ \\
\hline\hline
\end{tabular}
\normalsize
\end{table}

\FloatBarrier

\section{Conclusion and discussion}\label{sec4}

In this paper, we have studied linear scalar perturbations of KN black holes with quadratic CS and Maxwell couplings using hyperboloidal time-domain evolution and stationary axisymmetric zero modes.
The stationary calculation retains the full angular dependence and determines the global thresholds, which are compared with growth and decay in time-domain evolutions at representative parameters.
For $Q/M=0.4$, the pure-Maxwell case admits positive- and negative-coupling threshold branches with unequal magnitudes.
With equal dimensionless couplings, both threshold magnitudes decrease as spin increases, and $\lambda=-10$ gives a global onset at $a_{\rm crit}/M\simeq0.1558$.
This is well below the northern-polar local sign-change spin $a_{\rm loc}/M\simeq0.8$, since the odd CS source permits negative effective mass squared in the southern exterior at lower spins.
A local negative-mass region alone is therefore insufficient to determine the global onset.
For unequal couplings, the two real threshold branches merge as spin or CS coupling increases, with the stronger fixed CS coupling giving a merger at smaller spin.
At $a/M=0.9$ and $\hat\gamma=1$, the sampled signals decay between the thresholds $\eta=-82.2081$ and $-24.0552$ and grow outside them.

For the four axisymmetric initial multipoles $\ell_{\rm init}=0,1,2,3$ tested at $(a/M,Q/M,\hat\gamma,\eta)=(0.9,0.4,1,-20)$, mode mixing leads to late-time monopole dominance, with $P_0>0.99$ in the retained projection and nearly identical growth rates.
The selected $m=1,2$ waveforms decay over the simulated interval, although establishing stability of these sectors requires a separate spectral analysis.
Near marginality, the fitted rates depend on the initial data and fitting window, so they serve as dynamical checks of the stationary thresholds.

The present analysis can be extended to scalarized rotating charged black holes to examine how the CS and Maxwell couplings affect their existence and equatorial structure. Coupled perturbations, including nonaxisymmetric modes, and nonlinear evolutions could clarify whether unstable KN black holes settle into stable scalarized states. If such states exist, their quasinormal spectra and ringdown response could be compared with those of bald KN black holes at the same mass, angular momentum, and charge to identify signatures of scalar hair.

\section*{Acknowledgments}

This work is supported by the National Natural Science Foundation of China (Nos.12305064, 12665010, 12365009, 12405064 and 12565010), Jiangxi Provincial
Natural Science Foundation (Nos.20242BCE50055, 20262BAC240347, 20262BAC240348), National Research Foundation of Korea (NRF) grant funded by the Korea government(MSIT) (RS-2022-NR069013)
and Natural Science Basic Research Program of
Shaanxi Province (Program No. 2026JC-YBMS-0083).

\bibliographystyle{apsrev4-2}
\bibliography{references}

@article{Carter:1971zc,
  author  = {Carter, B.},
  journal = {Phys. Rev. Lett.},
  volume  = {26},
  pages   = {331},
  year    = {1971}
}

@article{Ruffini:1971bza,
  author  = {Ruffini, R. and Wheeler, J. A.},
  journal = {Phys. Today},
  volume  = {24},
  number  = {1},
  pages   = {30},
  year    = {1971}
}

@article{Bekenstein:1974sf,
  author  = {Bekenstein, J. D.},
  journal = {Annals Phys.},
  volume  = {82},
  pages   = {535},
  year    = {1974}
}

@article{Bekenstein:1975ts,
  author  = {Bekenstein, J. D.},
  journal = {Annals Phys.},
  volume  = {91},
  pages   = {75},
  year    = {1975}
}

@article{Bronnikov:1978mx,
  author  = {Bronnikov, K. A. and Kireev, Y. N.},
  journal = {Phys. Lett. A},
  volume  = {67},
  pages   = {95},
  year    = {1978}
}

@article{Sotiriou:2011dz,
  author  = {Sotiriou, T. P. and Faraoni, V.},
  journal = {Phys. Rev. Lett.},
  volume  = {108},
  pages   = {081103},
  year    = {2012}
}

@article{Herdeiro:2015waa,
  author  = {Herdeiro, C. A. R. and Radu, E.},
  journal = {Int. J. Mod. Phys. D},
  volume  = {24},
  number  = {09},
  pages   = {1542014},
  year    = {2015}
}

@article{Damour:1993hw,
  author  = {Damour, T. and Esposito-Farese, G.},
  journal = {Phys. Rev. Lett.},
  volume  = {70},
  pages   = {2220},
  year    = {1993}
}

@article{Kanti:1995vq,
  author  = {Kanti, P. and Mavromatos, N. E. and Rizos, J. and Tamvakis, K. and Winstanley, E.},
  journal = {Phys. Rev. D},
  volume  = {54},
  pages   = {5049},
  year    = {1996}
}

@article{Sotiriou:2013qea,
  author  = {Sotiriou, T. P. and Zhou, S. Y.},
  journal = {Phys. Rev. Lett.},
  volume  = {112},
  pages   = {251102},
  year    = {2014}
}

@article{Sotiriou:2014pfa,
  author  = {Sotiriou, T. P. and Zhou, S. Y.},
  journal = {Phys. Rev. D},
  volume  = {90},
  pages   = {124063},
  year    = {2014}
}

@article{Doneva:2017bvd,
  author  = {Doneva, D. D. and Yazadjiev, S. S.},
  journal = {Phys. Rev. Lett.},
  volume  = {120},
  number  = {13},
  pages   = {131103},
  year    = {2018}
}

@article{Silva:2017uqg,
  author  = {Silva, H. O. and Sakstein, J. and Gualtieri, L. and Sotiriou, T. P. and Berti, E.},
  journal = {Phys. Rev. Lett.},
  volume  = {120},
  number  = {13},
  pages   = {131104},
  year    = {2018}
}

@article{Antoniou:2017acq,
  author  = {Antoniou, G. and Bakopoulos, A. and Kanti, P.},
  journal = {Phys. Rev. Lett.},
  volume  = {120},
  number  = {13},
  pages   = {131102},
  year    = {2018}
}

@article{Minamitsuji:2018xde,
  author  = {Minamitsuji, M. and Ikeda, T.},
  journal = {Phys. Rev. D},
  volume  = {99},
  number  = {4},
  pages   = {044017},
  year    = {2019}
}

@article{Cunha:2019dwb,
  author  = {Cunha, P. V. P. and Herdeiro, C. A. R. and Radu, E.},
  journal = {Phys. Rev. Lett.},
  volume  = {123},
  number  = {1},
  pages   = {011101},
  year    = {2019}
}

@article{Collodel:2019kkx,
  author  = {Collodel, L. G. and Kleihaus, B. and Kunz, J. and Berti, E.},
  journal = {Class. Quant. Grav.},
  volume  = {37},
  number  = {7},
  pages   = {075018},
  year    = {2020}
}

@article{Dima:2020yac,
  author  = {Dima, A. and Barausse, E. and Franchini, N. and Sotiriou, T. P.},
  journal = {Phys. Rev. Lett.},
  volume  = {125},
  number  = {23},
  pages   = {231101},
  year    = {2020}
}

@article{Hod:2020jjy,
  author  = {Hod, S.},
  journal = {Phys. Rev. D},
  volume  = {102},
  number  = {8},
  pages   = {084060},
  year    = {2020}
}

@article{Zhang:2020pko,
  author  = {Zhang, S. J. and Wang, B. and Wang, A. and Saavedra, J. F.},
  journal = {Phys. Rev. D},
  volume  = {102},
  number  = {12},
  pages   = {124056},
  year    = {2020}
}

@article{Doneva:2020nbb,
  author  = {Doneva, D. D. and Collodel, L. G. and Krueger, C. J. and Yazadjiev, S. S.},
  journal = {Phys. Rev. D},
  volume  = {102},
  number  = {10},
  pages   = {104027},
  year    = {2020}
}

@article{Doneva:2020kfv,
  author  = {Doneva, D. D. and Collodel, L. G. and Krueger, C. J. and Yazadjiev, S. S.},
  journal = {Eur. Phys. J. C},
  volume  = {80},
  number  = {12},
  pages   = {1205},
  year    = {2020}
}

@article{Herdeiro:2020wei,
  author  = {Herdeiro, C. A. R. and Radu, E. and Silva, H. O. and Sotiriou, T. P. and Yunes, N.},
  journal = {Phys. Rev. Lett.},
  volume  = {126},
  number  = {1},
  pages   = {011103},
  year    = {2021}
}

@article{Berti:2020kgk,
  author  = {Berti, E. and Collodel, L. G. and Kleihaus, B. and Kunz, J.},
  journal = {Phys. Rev. Lett.},
  volume  = {126},
  number  = {1},
  pages   = {011104},
  year    = {2021}
}

@article{Alexander:2009tp,
  author  = {Alexander, S. and Yunes, N.},
  journal = {Phys. Rept.},
  volume  = {480},
  pages   = {1},
  year    = {2009}
}

@article{Yunes:2009hc,
  author  = {Yunes, N. and Pretorius, F.},
  journal = {Phys. Rev. D},
  volume  = {79},
  pages   = {084043},
  year    = {2009}
}

@article{Myung:2020etf,
  author  = {Myung, Y. S. and Zou, D. C.},
  journal = {Phys. Lett. B},
  volume  = {814},
  pages   = {136081},
  year    = {2021}
}

@article{Zou:2021ybk,
  author  = {Zou, D. C. and Myung, Y. S.},
  journal = {Phys. Lett. B},
  volume  = {820},
  pages   = {136545},
  year    = {2021}
}

@article{Gao:2018cs,
  author        = {Gao, Yuan-Xing and Huang, Yang and Liu, Dao-Jun},
  title         = {Scalar perturbations on the background of Kerr black holes in the quadratic dynamical Chern--Simons gravity},
  journal       = {Phys. Rev. D},
  volume        = {99},
  pages         = {044020},
  year          = {2019},
  doi           = {10.1103/PhysRevD.99.044020},
  eprint        = {1808.01433},
  archivePrefix = {arXiv},
  primaryClass  = {gr-qc}
}

@article{Herdeiro:2018wub,
  author  = {Herdeiro, C. A. R. and Radu, E. and Sanchis-Gual, N. and Font, J. A.},
  journal = {Phys. Rev. Lett.},
  volume  = {121},
  number  = {10},
  pages   = {101102},
  year    = {2018}
}

@article{Fernandes:2019rez,
  author  = {Fernandes, P. G. S. and Herdeiro, C. A. R. and Pombo, A. M. and Radu, E. and Sanchis-Gual, N.},
  journal = {Class. Quant. Grav.},
  volume  = {36},
  number  = {13},
  pages   = {134002},
  year    = {2019}
}

@article{Astefanesei:2019pfq,
  author  = {Astefanesei, D. and Herdeiro, C. and Pombo, A. and Radu, E.},
  journal = {JHEP},
  volume  = {2019},
  number  = {10},
  pages   = {078},
  year    = {2019}
}

@article{Myung:2018vug,
  author  = {Myung, Y. S. and Zou, D. C.},
  journal = {Eur. Phys. J. C},
  volume  = {79},
  number  = {3},
  pages   = {273},
  year    = {2019}
}

@article{Lai:2022spn,
  author  = {Lai, M. Y. and Myung, Y. S. and Yue, R. H. and Zou, D. C.},
  journal = {Phys. Rev. D},
  volume  = {106},
  number  = {4},
  pages   = {044045},
  year    = {2022}
}

@article{Lai:2022cjs,
  author  = {Lai, M. Y. and Myung, Y. S. and Yue, R. H. and Zou, D. C.},
  journal = {Phys. Rev. D},
  volume  = {106},
  number  = {8},
  pages   = {084043},
  year    = {2022}
}

@article{Zenginoglu:2011zz,
  author  = {Zenginoglu, A. and Khanna, G.},
  journal = {Phys. Rev. X},
  volume  = {1},
  pages   = {021017},
  year    = {2011}
}

@article{Thuestad:2017ngu,
  author  = {Thuestad, I. and Khanna, G. and Price, R. H.},
  journal = {Phys. Rev. D},
  volume  = {96},
  number  = {2},
  pages   = {024020},
  year    = {2017}
}

@article{Zenginoglu:2007jw,
  author  = {Zenginoglu, A.},
  journal = {Class. Quant. Grav.},
  volume  = {25},
  pages   = {145002},
  year    = {2008}
}

@article{Racz:2011qu,
  author  = {Racz, I. and Toth, G. Z.},
  journal = {Class. Quant. Grav.},
  volume  = {28},
  pages   = {195003},
  year    = {2011}
}

@article{Harms:2013ib,
  author  = {Harms, E. and Bernuzzi, S. and Bruegmann, B.},
  journal = {Class. Quant. Grav.},
  volume  = {30},
  pages   = {115013},
  year    = {2013}
}

@article{Harms:2014dqa,
  author  = {Harms, E. and Bernuzzi, S. and Nagar, A. and Zenginoglu, A.},
  journal = {Class. Quant. Grav.},
  volume  = {31},
  number  = {24},
  pages   = {245004},
  year    = {2014}
}

@article{Zenginoglu:2010cq,
  author  = {Zenginoglu, A.},
  journal = {J. Comput. Phys.},
  volume  = {230},
  pages   = {2286},
  year    = {2011}
}

@book{chandra,
  author    = {Chandrasekhar, S.},
  title     = {The Mathematical Theory of Black Holes},
  publisher = {Oxford University Press},
  address   = {New York},
  year      = {1983}
}

@article{Fan:2023jhi,
  author        = {Fan, K.-H. and Myung, Y. S. and Zou, D.-C. and Lai, M.-Y.},
  title         = {Scalarization of Kerr--Newman black holes in the Einstein--Chern--Simons-scalar theory},
  journal       = {Eur. Phys. J. C},
  volume        = {84},
  number        = {7},
  pages         = {679},
  year          = {2024},
  doi           = {10.1140/epjc/s10052-024-13030-y},
  eprint        = {2401.00144},
  archivePrefix = {arXiv},
  primaryClass  = {gr-qc}
}

@article{Guo:2023mda,
  author        = {Guo, G. and Wang, P. and Wu, H. and Yang, H.},
  title         = {Scalarized Kerr--Newman black holes},
  journal       = {JHEP},
  volume        = {2023},
  number        = {10},
  pages         = {076},
  year          = {2023},
  doi           = {10.1007/JHEP10(2023)076},
  eprint        = {2307.12210},
  archivePrefix = {arXiv},
  primaryClass  = {gr-qc}
}

@article{Hod:2023jdc,
  author        = {Hod, S.},
  title         = {Analytic study of the Maxwell electromagnetic invariant in spinning and charged Kerr--Newman black-hole spacetimes},
  journal       = {JHEP},
  volume        = {2023},
  number        = {09},
  pages         = {140},
  year          = {2023},
  doi           = {10.1007/JHEP09(2023)140},
  eprint        = {2308.12990},
  archivePrefix = {arXiv},
  primaryClass  = {gr-qc}
}

@article{Hod:2022txa,
  author        = {Hod, S.},
  title         = {Spin-charge induced scalarization of Kerr--Newman black-hole spacetimes},
  journal       = {JHEP},
  volume        = {2022},
  number        = {08},
  pages         = {272},
  year          = {2022},
  doi           = {10.1007/JHEP08(2022)272},
  eprint        = {2206.12074},
  archivePrefix = {arXiv},
  primaryClass  = {gr-qc}
}

@book{Roache:1998,
  author    = {Roache, P. J.},
  title     = {Verification and Validation in Computational Science and Engineering},
  publisher = {Hermosa Publishers},
  address   = {Albuquerque},
  year      = {1998}
}

@book{Press:2007,
  author    = {Press, W. H. and Teukolsky, S. A. and Vetterling, W. T. and Flannery, B. P.},
  title     = {Numerical Recipes: The Art of Scientific Computing},
  edition   = {3},
  publisher = {Cambridge University Press},
  address   = {Cambridge},
  year      = {2007}
}

@article{Bhagwat:2017tkm,
    author = "Bhagwat, Swetha and Okounkova, Maria and Ballmer, Stefan W. and Brown, Duncan A. and Giesler, Matthew and Scheel, Mark A. and Teukolsky, Saul A.",
    title = "{On choosing the start time of binary black hole ringdowns}",
    journal = "Phys. Rev. D",
    volume = "97",
    number = "10",
    pages = "104065",
    year = "2018",
    eprint = "1711.00926",
    archivePrefix = "arXiv",
    primaryClass = "gr-qc",
    doi = "10.1103/PhysRevD.97.104065"
}

@article{Giesler:2019uxc,
    author = "Giesler, Matthew and Isi, Maximiliano and Scheel, Mark A. and Teukolsky, Saul A.",
    title = "{Black Hole Ringdown: The Importance of Overtones}",
    journal = "Phys. Rev. X",
    volume = "9",
    number = "4",
    pages = "041060",
    year = "2019",
    eprint = "1903.08284",
    archivePrefix = "arXiv",
    primaryClass = "gr-qc",
    doi = "10.1103/PhysRevX.9.041060"
}

@article{Witek:2018dmd,
    author = "Witek, Helvi and Gualtieri, Leonardo and Pani, Paolo and Sotiriou, Thomas P.",
    title = "{Black holes and neutron stars in scalar-tensor theories of gravity: Nonlinear dynamics in spherical symmetry}",
    journal = "Phys. Rev. D",
    volume = "99",
    number = "6",
    pages = "064035",
    year = "2019",
    eprint = "1810.05177",
    archivePrefix = "arXiv",
    primaryClass = "gr-qc",
    doi = "10.1103/PhysRevD.99.064035"
}

@book{Baumgarte:2010,
    author = "Baumgarte, Thomas W. and Shapiro, Stuart L.",
    title = "{Numerical Relativity: Solving Einstein's Equations on the Computer}",
    publisher = "Cambridge University Press",
    year = "2010",
    doi = "10.1017/CBO9781139193344"
}

@article{Sperhake:2014wpa,
    author = "Sperhake, Ulrich",
    title = "{The numerical relativity breakthrough for binary black holes}",
    journal = "Living Rev. Relativ.",
    volume = "18",
    pages = "1",
    year = "2015",
    eprint = "1411.3997",
    archivePrefix = "arXiv",
    primaryClass = "gr-qc",
    doi = "10.1007/lrr-2015-1"
}
\end{document}